\documentclass[twocolumn]{aastex701}
\usepackage{subcaption}
\usepackage{amsmath}
\usepackage{hyperref}

\begin{document}

\title[Polarimetric study of AFGL 6366S]{Probing the magnetic fields and dust properties in the young embedded star-forming region AFGL 6366S using Near Infrared and Optical linear polarimetry}

\author[orcid=0009-0005-2615-4547,sname='Biswas']{Samrat Biswas}
\affiliation{Department of Physics, Gauhati University, Gopinath Bordoloi Nagar, Jalukbari, Guwahati 781014 ,Assam,India}
\email[show]{samrat@gauhati.ac.in}  

\author[orcid= 0000-0002-3448-8150,sname='Medhi']{Biman J. Medhi} 
\affiliation{Department of Physics, Gauhati University, Gopinath Bordoloi Nagar, Jalukbari, Guwahati 781014 ,Assam,India}
\affiliation{Aryabhatta Research Institute of Observational Sciences, Manora Peak, Nainital - 263 129, India}
\email[show]{bimanjmedhi@gmail.com}  

\author[orcid= 0000-0002-6510-0681,sname='Tamura']{Motohide Tamura}
\affiliation{Department of Astronomy, Graduate School of Science, The University of Tokyo, 7-3-1 Hongo, Bunkyo-ku, Tokyo 113-0033, Japan}
\affiliation{National Astronomical Observatory of Japan, National Institutes of Natural Sciences, Osawa, Mitaka, Tokyo 181-8588, Japan}
\email{motohide.tamura@nao.ac.jp} 

\author[orcid=0000-0003-0262-7264,sname='Das']{H.S Das}
\affiliation{Department of Physics, Assam University, Silchar 788011, India}
\email{hsdas13@gmail.com} 

\author[orcid= 0000-0003-2815-7774,sname='Kwon']{Jungmi Kwon}
\affiliation{Department of Astronomy, Graduate School of Science, The University of Tokyo, 7-3-1 Hongo, Bunkyo-ku, Tokyo 113-0033, Japan}
\email{kwon.jungmi@astron.s.u-tokyo.ac.jp}

\begin{abstract}
We present Near-Infrared (NIR) and Optical linear polarimetry towards the partially embedded cluster AFGL 6366S. The polarization ranges from 0.44-10.3 per cent in NIR and 0.16-11.22 per cent in Optical bands. The position angle spans $1^\circ - 179^\circ$ in both NIR and Optical bands. About 22 stars exhibit intrinsic polarization signatures. A polarization hole is evident towards the densest ( $\sim~3.4 \times 10^{23}\,\mathrm{cm}^{-2}$) and warmest ($\sim~ 28.8\,\mathrm{K}$) central cluster region. It is attributable to depolarization induced by Radiative Torque Disruption (RAT-D) of large grains and a modest contribution from magnetic-field tangling. The local magnetic field towards the cluster’s central region is significantly misaligned with both the large-scale Galactic field and the long axis of the filament present in the region. The field morphology wraps around two dense molecular clumps of radii 0.34 pc and 0.22 pc and $\mathrm N(\mathrm H_2)$ = $(7.9 \pm 1.1) \times 10^{22}$ cm$^{-2}$ and $(4.3 \pm 0.5) \times 10^{22}$ cm$^{-2}$, respectively. The clumps are embedded in the filamentary structure and represent locally accelerated stages of mass accumulation. Gravitationally driven mass flows, largely perpendicular to the local magnetic field, produce a U-shaped field curvature across the filament axis. The plane-of-sky magnetic field strengths towards the two clumps are \( 447.91 \pm 83.81~\mu\mathrm{G}\) and \(396.66 \pm 73.64~\mu\mathrm{G}\). The corresponding mass-to-flux ratios (\( \lambda \sim~ 1.34\) and \(0.82\)) indicate that one clump is magnetically supercritical and the other is subcritical. The Alfv\'en Mach numbers ($\mathcal{M}_A$) $\sim$ 0.395 and 0.393 indicate that both the clumps are in sub-Alfv\'enic state.
\end{abstract}


\section{Introduction}
\begingroup
\renewcommand\thefootnote{}\footnote{"Herschel is an ESA space observatory \citep{Pilbratt_2010} with science instruments provided by European-led Principal Investigator consortia and with important participation from NASA"}
\endgroup
Interstellar dust and magnetic fields are two fundamental components that shape the physical and dynamical conditions of the interstellar medium (ISM) \citep{1987MNRAS.224..413T, 1993prpl.conf..327M, 2007MNRAS.377...77P,MRK_2019}. Dust, in particular, governs the radiative and chemical processes by shielding molecules from energetic radiation. This facilitates molecular cloud cooling and the subsequent collapse \citep{Draine_2003, 2024ApJ...975..284S, 2010MNRAS.403.1894D, 2022FrASS...9.8217T}. On the other hand, magnetic fields regulate gas dynamics and provide support against the gravitational collapse \citep{1993prpl.conf..327M, 2004mim..proc..123C,2024MNRAS.528.1460R}. Even with extensive observational, theoretical, and numerical efforts, however, the coupling between these components and its role in regulating star formation remains incompletely understood \citep{MRK_2019}.

 Typically, the interstellar dust grains are non-spherical in shape. It is due to the processes such as coagulation, mantle accretion and fragmentation in interstellar shocks \citep{Hoang_2022}. When these asymmetric grains become aligned with the local magnetic field, they differentially absorb components of the incident starlight. As a result, it produces a net linear polarization. Stellar polarimetry therefore provides a robust means of tracing both dust properties and magnetic-field structure in the ISM \citep[e.g.][]{spitzer_1951, davis_1951, serkowski_1973, medhi_2007, Medhi_2010, singh_2020, Biswas_2024}. 
 
An integrated analysis of background-star polarization along with the region’s column density and dust temperature provide a sensitive tracer of the conditions governing grain alignment \citep[e.g.][]{2025ApJ...981..128P}. Several mechanisms for grain alignment have been proposed in the literature \citep{Lazarian_1997, Lazarian_2003, Voshchinnikov_2012}. Among these, the radiative torque alignment (RAT) framework currently has the strongest observational and theoretical support. In this model, anisotropic radiation aligns the grain’s shortest axis parallel to the local magnetic-field direction \citep{1996ApJ...470..551D, 2007MNRAS.378..910L}.

 Regions with strong magnetic regulation show smoothly ordered polarization angles. The turbulence introduces measurable angular dispersion \citep{2009ApJ...696..567H}. Thus, polarization studies serve as a powerful diagnostic of the role of magnetic-turbulent interactions in shaping the interstellar clouds \citep{1976ApJ...210..326M, 1978PASJ...30..671N, shu1987ARA&A..25...23S, 2004mim..proc..123C, Girart2009Sci...324.1408G}.

Nonetheless, the effectiveness of stellar polarimetry strongly depends on the choice of observational wavelength. Optical ($\sim$ 400 - 700 nm) polarimetry is particularly well-suited for regions with relatively low dust extinction ($A_V \lesssim 3 \rm~ mag$) \citep[e.g.,][]{Biswas_2024}. However, in dense, partially embedded ($A_V \sim 3-30 \rm~mag$) regions Near-Infrared (NIR $\sim$ 1 - 2 $\mu$m) polarimetry offers a more viable alternative \citep[e.g.,][]{1987MNRAS.224..413T}. For heavily obscured environments ($A_V > 30 \rm~mag$)\footnote{The extinction ranges quoted in parentheses follow the classification of \citet{Kandori_2018}.}, Far-Infrared (FIR $\sim$ 50 - 300 $\mu$m)/Submillimeter(Submm $\sim$ 300 - 850 $\mu$m)/ Millimeter(mm $\sim$ 1 - 3 mm) 'thermal' polarimetry provide the most effective probes \citep[e.g.,][]{1999sf99.proc..212T}.

In this work, we combine NIR and Optical polarimetric observations with multi-wavelength archival data to investigate the region towards the young ($\sim$ 3 Myr) star cluster AFGL 6366S \citep[Right Ascension (R.A.): $06^h08^m40^s$, Declination (Dec.): $+21^{\circ}31'08''$;][]{shimoikura_2013, 2025MNRAS.541.1557B}. It is a high-mass star-forming region and lies adjacent to the H II region Sh2-247 within the Gemini OB1 cloud. The region hosts a Class II 6.7 GHz CH$_3$OH maser, 22 GHz H$_2$O maser, and OH masers at 1665–1720 MHz \citep{1979A&AS...37....1T,1989A&A...221..295K,Szymczak2018MNRAS.474..219S,2020MNRAS.499.1441D}.  A CO (J = 2-1) bipolar outflow, weak 3.6 cm radio continuum emission and and millimetre continuum emission were also reported in this region \citep{Snell1988ApJ...325..853S, 1994ApJS...91..659K, 2004A&A...426..503W, 2010ApJS..188..123R}.

The first NIR polarimetric study of AFGL 6366S was conducted by \cite{2000ApJ...542..392Y} using the 1.88 m telescope at Okayama Astrophysical Observatory, Japan. Their work primarily involved detection of deeply embedded sources in the region through inspection of polarimetric patterns. More recently, \cite{Kwon2018AJ....156....1K} presented NIR linear and circular imaging polarimetry of this region using the InfraRed Survey Facility (IRSF) 1.4 m telescope in South Africa. Their observations revealed two bright polarized nebulosities. They also reported the first detection of circular polarization in this region, which is characterized by an asymmetric quadrupolar pattern across the cluster.

The present work builds upon our earlier study of the fundamental, structural and dynamical properties of AFGL 6366S \citep[][Paper I hereafter]{2025MNRAS.541.1557B}. This cluster still remains partially embedded in its natal dust and molecular material. The member stars exhibit a fractal spatial distribution, suggesting formation in a turbulent environment with signatures of hierarchical star formation. These properties make AFGL 6366S an ideal target for investigating the relative importance of dust and magnetic fields during early phases of star formation.This study aims to provide a comprehensive understanding of the physical conditions governing the formation and evolution of this young cluster. 

The paper is organized as following:  Section \ref{sec:archival_data} describes the archival data used. Sec. \ref{sec:obs_data_red} presents the observations and data reduction. The analysis and results are presented in sec. \ref{sec:result_1}. The discussions and the summary are  presented in Sec. \ref{sec:discuss} and \ref{sec:summ}, respectively.

\section{Archival data}
\label{sec:archival_data}

\subsection{Planck data}

The \textit{Planck} satellite mapped the linear polarization of the entire sky across seven frequency bands. It ranges from 30 to 353~GHz \citep{2014A&A...571A...1P}. The data are publicly available via the \href{https://pla.esac.esa.int}{Planck Legacy Archive}.
In this study, we use the polarized thermal dust emission from the \textit{Planck}-HFI 353~GHz channel to investigate the large scale plane-of-sky (POS) magnetic field towards the AFGL 6366S region. The \textit{Planck} observations provide all-sky maps of the Stokes parameters $I$, $Q$ and $U$. From these Stokes parameters, the polarization fraction ($p$), polarization angle ($\psi$) and the POS magnetic field orientation ($\theta_B$) are computed using the following relations:

\begin{equation}
	p = \frac{\sqrt{Q^{2} + U^{2}}}{I},
\end{equation}

\begin{equation}
	\psi = \frac{1}{2} \arctan(-U, Q),
	\label{Eq:planck_psi}
\end{equation}

\begin{equation}
	\theta_B = \psi + \frac{\pi}{2}.
	\label{Eq:planck_theta}
\end{equation}

In Eq. \ref{Eq:planck_psi}, the two-argument function $\arctan(-U, Q)$ ensures the correct quadrant for the polarization angle. The negative sign adjusts the \textit{Planck} data from the \texttt{HEALPix} convention to the IAU standard. According to the IAU standard, the polarization angles are measured counterclockwise from the Galactic north toward the east. Further methodological details can be found in \citet{2015A&A...576A.104P}. 


\subsection{2MASS NIR $JHK_{s}$ Data}
\label{Sec:2mass_data}
To identify intrinsically polarized sources in our NIR polarimetric sample, we obtained their corresponding $JHK_s$ photometric data from the Two Micron All Sky Survey point source catalog \footnote{Accessible at \url{https://vizier.cds.unistra.fr/viz-bin/VizieR-3?-source=II/246/out}} \citep[2MASS PSC;][]{Cutri_2003, Skrutskie_2006}. These were then used to construct the polarization-colour diagram (PCD; Fig. \ref{Fig:nir_in_pol}). The polarimetric data was cross-matched with the 2MASS PSC using the \href{http://cdsxmatch.u-strasbg.fr/}{CDS XMatch} service. A search radius of $2''$ around each target was adopted for the cross match.

For a reliable selection, we retained only the objects with high-quality photometry, i.e.\ sources with a 2MASS quality flag of \textit{AAA}\footnote{Each character corresponds to the photometric quality in the $J$, $H$, and $K_s$ bands, respectively. The flag ``\textit{AAA}'' denotes a signal-to-noise ratio $\ge 10$ and photometric uncertainty $\le 0.10857$ in all three bands \citep{Cutri_2003}.} and statistically significant polarization detections i.e, $P/e_p > 3$, where $P$ is the measured degree of polarization and $e_p$ is its associated uncertainty. 

After applying these criteria, a total of 53 out of the 64 NIR polarimetric sources were retained for the construction of the PCD. The selected sources were within the magnitude limits of $J \le 17.088$ mag, $H \le 14.736$ mag, and $K_s \le 13.666$ mag.

\subsection{\textit{Herschel} Data}

To investigate the distribution of dense dust structures in the AFGL 6366S region, we made use of the \textit{Herschel} Spectral and Photometric Imaging Receiver (SPIRE) observations. It provides high–quality imaging at 250, 350, and 500~$\mu$m \citep{Pilbratt_2010}. The dust continuum emission towards our target region was extracted from the SPIRE 500~$\mu$m map. The field of view (FOV) was centred on AFGL 6366S cluster. It covered an area of $50 \times 50$ arcmin$^{2}$. This field size was chosen to ensure complete coverage of the large-scale dense dust structures surrounding the cluster. These included the extended filaments and clump complexes. This wide coverage also allowed us to examine how these dust structures influence the large-scale magnetic-field orientation.

All data products were retrieved from the \href{https://archives.esac.esa.int/hsa/whsa/}{Herschel Science Archive}.

\section{Observation and data reduction}
\label{sec:obs_data_red}


\subsection{NIR polarimetric data from SIRPOL}

The NIR imaging polarimetric observations towards AFGL 6366S were obtained for 64 stars. The observed field is shown as a red square in Fig.~\ref{Fig: pol}. The observations were performed using SIRPOL \citep{2006SPIE.6269E..51K}, the polarimetric mode of the Simultaneous 3-color (\textit{JHK$_s$}) InfraRed Imager for Unbiased Survey (SIRIUS) camera, mounted on the 1.4 m IRSF telescope at SAAO, Sutherland. It is a single-beam polarimeter consisting of an achromatic half-wave plate (1–2.5 $\mu$m) with a rotator and a high-efficiency wire-grid polarizer placed upstream of the camera. A summary of the observational details and the key camera specifications are provided in Table~\ref{tab:obs_summary_nir}.

\begin{table}
\setlength{\tabcolsep}{2pt}
\centering
\caption{Summary of the NIR imaging polarimetric observations and the SIRIUS camera specifications.}
\label{tab:obs_summary_nir}
\begin{tabular}{ll}
\hline
Parameter & Description \\
\hline
\multicolumn{2}{c}{Observation details} \\
\hline
Observing Telescope & 1.4m IRSF telescope \\
Observing date & 20 February 2007 \\
Observation Field center & R.A. $06^{\text{h}}\,08^{\text{m}}\,40.9^{\text{s}}$ \\
 & Dec. $+21^{\circ}\,30^{\text{m}}\,59.9^{\text{s}}$ \\
Filters (central wavelengths) & \textit{J} (1.25 $\mu$m), \textit{H} (1.63 $\mu$m),\\
& \textit{K$_s$} (2.14 $\mu$m) \\
Exposure time (per frame)$^{(*)}$ & $\sim$ ~10 s \\
FWHM of stellar images & $\sim$ 3.5 - 3.8 pixels\\
\hline
\multicolumn{2}{c}{SIRIUS Camera Specifications} \\
\hline
Detector type & Three $1024 \times 1024$ \\
& HgCdTe arrays \\
Pixel scale & $0.453$ arcsec pixel$^{-1}$ \\
Field of view & $7.7 \times 7.7$ arcmin$^2$ \\
Readout noise ($e^-$) & 19 (\textit{J}), 25 (\textit{H}), 35 (\textit{K$_s$}) \\
Gain ($e^-$/ADU) & 5.1 (\textit{J}), 5.0 (\textit{H}), 5.3 (\textit{K$_s$}) \\
Limiting sensitivity & $J = 19.2$ mag \\
\hline
\end{tabular}
\setlength{\leftskip}{-25pt}\\
\setlength{\leftskip}{0pt}
\small{Notes: $^{(*)}$Each frame corresponds to one dither position at a given HWP angle and filter.}
\end{table}

Linear polarimetric observations of the target stars were conducted by rotating the half-wave plate (HWP) through four angular positions: $0^{\circ}$, $22.5^{\circ}$, $45^{\circ}$, and $67.5^{\circ}$. For each HWP position, measurements were taken at 10 dithered locations, with an exposure time of 10$~$s per frame. This observation sequence was repeated 8 times for both the target and background sky frames to improve the signal-to-noise ratio (SNR). 

The data reduction was carried out using the SIRPOL pipeline package that uses Image Reduction and Analysis facility (IRAF)\footnote{"IRAF is distributed by National Optical Astronomical Observatories, USA"} and routines written in C language. The standard reduction steps included flat-field correction, sky subtraction and combining of the dithered frames. The Stokes' parameters $I, Q$ and $U$ were derived from a set of exposures obtained at the four HWP angles at the same dithered positions, using the expressions in \citet{2006SPIE.6269E..51K}.




The degree of polarization ($P$) and position angle ($\theta$) were estimated using the relations \citep{2006SPIE.6269E..51K}: 

\begin{equation}
P = \frac{\sqrt{Q^2 + U^2}}{I} ~~~ \text{and} ~~~ \theta = \frac{1}{2} tan^{-1} \left(\frac{U}{Q}\right)
\end{equation}

The uncertainty in the polarization ($e_P$) were estimated as described in \cite{2016ApJS..222....2K}.
The error in $\theta$ is estimated using the standard formula \citep{2009AIPC.1158..111H}: 

\begin{equation}
e_\theta = 28.65^\circ \times \frac{e_P}{P} 
\end{equation}

Further details of the SIRPOL instrument and its capabilities are discussed in \cite{2006SPIE.6269E..51K}

\subsection{Optical polarimetric data from AIMPOL}
\label{Sec: Optical_polarimetry}

\begin{table}
\setlength{\tabcolsep}{2pt}
\centering
\caption{Summary of the Optical imaging polarimetric observations and the CCD specifications.}
\label{tab:obs_summary_opt}
\begin{tabular}{ll}
\hline
Parameter & Description \\
\hline
\multicolumn{2}{c}{Observation details} \\
\hline
Observing Telescope & 104cm Sampurnand telescope \\
Observing date & 24th and 25th November 2011 \\
Observation Field center & R.A. $06^{\text{h}}\,08^{\text{m}}\,40.8^{\text{s}}$ \\
 & Dec. $+21^{\circ}\,30^{\text{m}}\,43^{\text{s}}$ \\
Filters (central wavelengths) & \textit{B} (0.44 $\mu$m), \textit{V} (0.55 $\mu$m),\\
& \textit{R$_c$} (0.66 $\mu$m), \textit{I$_c$} (0.80 $\mu$m) \\
Exposure time (per frame) & 300s (\textit{B}), 280s (\textit{V}),\\
& 240s (\textit{R$_c$}), 200s (\textit{I$_c$}) \\
FWHM of stellar images & $\sim$ 2 - 3 pixels\\
\hline
\multicolumn{2}{c}{CCD Specifications} \\
\hline
Detector type & TK $1024 \times 1024$ pixel$^2$\\
Pixel scale & $1.73$ arcsec pixel$^{-1}$ \\
Field of view & 8 arcmin diameter \\
Readout noise ($e^-$) & 7.0 \\
Gain ($e^-$/ADU) & 11.98 \\
Limiting sensitivity & $V = 18$ mag \\
\hline
\end{tabular}
\end{table}

\begin{table*}
\centering
\caption{Observed polarized and unpolarized standard stars. $P$ $\pm$ $e_p$ and $\theta$ $\pm$ $e_\theta$ denote the degree and position angle of polarization with their corresponding uncertainties, respectively. $q$ and $u$ are the normalized Stokes parameters.}\label{std_obs}
\begin{tabular}{lllllll}
\hline
\multicolumn{5}{c}{Polarized Standard}&\multicolumn{2}{c}{Unpolarized Standard}\\
\hline
Filter&$P\pm e_p$ (per cent) &  $\theta \pm e_\theta (^\circ)$ &  $P\pm e_p$ (per cent) & $\theta \pm e_\theta (^\circ)$ &\ \ \ \ $q$ (per cent) &\ \ \ \ \ $u$ (per cent) \\
\hline
& \multicolumn{2}{c}{\citep{schmidt_1992}}&\multicolumn{2}{c}{This work}&\multicolumn{2}{c}{This work}\\
\hline
\multicolumn{5}{c}{\underline{Hiltner-960}}                            &\multicolumn{2}{c}{\underline{HD21447}} \\
B & $5.72\pm0.06$ & $55.06\pm 0.31$  & $5.62\pm 0.20$&$  54.65 \pm 1.04$ & \ \ \ \ \ 0.019  &\ \ \ \ \ \ 0.013  \\
V & $5.66\pm0.02$ & $54.79\pm 0.11$  & $5.70\pm 0.14$&$  53.37 \pm 0.08$ & \ \ \ \ \ 0.038  &\ \ \ \ -\ 0.030 \\
$R_c$ & $5.21\pm0.03$ & $54.54\pm 0.16$  & $5.20\pm 0.06$&$  54.80 \pm 0.38$ & \ \ \ -\ 0.036  &\ \ \ \ -\ 0.038 \\
\multicolumn{5}{c}{\underline{HD 204827}}                              &\multicolumn{2}{c}{\underline{HD12021}} \\
B & $5.65\pm0.02$ & $58.20\pm 0.11$  & $5.72\pm 0.09$&$  58.60 \pm 0.49$ & \ \ \ -\ 0.107  & \ \ \ \ \ \ 0.072 \\
V & $5.32\pm0.02$ & $58.73\pm 0.08$  & $5.35\pm 0.03$&$  60.10 \pm 0.20$ & \ \ \ \ \ 0.043 & \ \ \ \ -\ 0.044 \\
$R_c$ & $4.89\pm0.03$ & $59.10\pm 0.17$  & $4.91\pm 0.20$&$  58.89 \pm 1.20$ & \ \ \ \ \ 0.021 & \ \ \ \ \ \ 0.032 \\
\multicolumn{5}{c}{\underline{BD+64$^{\circ}$106}}                     &\multicolumn{2}{c}{\underline{HD14069}} \\
B & $5.51\pm0.09$ & $97.15\pm 0.47$  & $5.46\pm 0.10$&$  99.40 \pm 0.50$ &\ \ \ \ \ 0.138 &\ \ \ \ -\ 0.011 \\
V & $5.69\pm0.04$ & $96.63\pm 0.18$  & $5.48\pm 0.11$&$  97.09 \pm 0.12$ &\ \ \ \ \ 0.022 &\ \ \ \ \ \  0.019 \\
$R_c$ & $5.15\pm0.10$ & $96.74\pm 0.54$  & $5.20\pm 0.02$&$  97.35 \pm 0.18$ &\ \ \ \ \ 0.011 &\ \ \ \ -\ 0.014 \\
\multicolumn{5}{c}{\underline{HD 19820}}                               &\multicolumn{2}{c}{\underline{G191B2B}} \\
B & $4.70\pm0.04$ & $115.70\pm 0.22$ & $4.81\pm 0.20$&$ 113.49 \pm 0.19$ &\ \ \ \ \ 0.073 &\ \ \ \ -\ 0.058 \\
V & $4.79\pm0.03$ & $114.93\pm 0.17$ & $4.91\pm 0.10$&$ 114.55 \pm 0.20$ &\ \ \ -\ 0.021 &\ \ \ \ -\ 0.042 \\
$R_c$ & $4.53\pm0.03$ & $114.46\pm 0.17$ & $4.70\pm 0.13$&$ 113.88 \pm 0.21$ &\ \ \ -\ 0.037 &\ \ \ \ \ \ 0.028 \\
\multicolumn{5}{c}{\underline{HD236633}}                               &\multicolumn{2}{c}{\underline{}} \\
$I_c$ & $4.81\pm0.04$ & $93.14\pm 0.21$ & $4.56\pm 0.10$&$ 93.39 \pm 0.67$ &\ \ \ \ \ ~~~~- &\ \ \ \ \ ~~~~- \\
\multicolumn{5}{c}{\underline{HD7927}}                               &\multicolumn{2}{c}{\underline{}} \\
$I_c$ & $2.78\pm0.03$ & $90.10\pm 0.31$ & $2.65\pm 0.041$&$ 89.86 \pm 0.44$ &\ \ \ \ \ ~~~~- &\ \ \ \ \ ~~~~- \\
\hline
\end{tabular}
\setlength{\leftskip}{-30pt}\\
\end{table*}

 Optical imaging polarimetric observations of a subset of 31 stars, among the 64 previously observed with SIRPOL, were carried out using the ARIES Imaging Polarimeter (AIMPOL; \citealt{Rj_2004}). AIMPOL is mounted at the back-end of the  104-cm Sampurnanand f/13 Cassegrain telescope of ARIES, Nainital. The observation field is almost coincident with that of the NIR polarimetric observation. It is shown in Fig.~\ref{Fig: pol} using the larger blue circle. A summary of the observational details and the specifications of the CCD camera used for imaging, are provided in Table~\ref{tab:obs_summary_opt}.

A comprehensive account of the AIMPOL instrument design and the data reduction methodology is available in \cite{Biswas_2024}.

To check the calibration of the position angle and to estimate the instrumental polarization, observations were also carried out for six polarized and four unpolarized standard stars listed in \cite{schmidt_1992}. The measured $P$ and $\theta$ for the polarized standard stars and their corresponding values from \cite{schmidt_1992} are presented in Table \ref{std_obs}. They are consistent within the quoted uncertainties. For the unpolarized standards, the measured normalized Stokes' parameters $q$ (per cent) and $u$ (per cent) are also provided in Table \ref{std_obs}. From these, the mean instrumental polarization was estimated to be $\sim  0.03$ per cent.   

\section{Analysis and Results}
\label{sec:result_1}

\begin{table*}\hspace{-20cm}
\setlength{\tabcolsep}{15pt} 
\centering
\caption{Observed $J$, $H$, and $K_s$ polarization measurements for 64 stars in AFGL~6366S. ID denotes the stellar identification number. R.A. and Dec. are the Right Ascension and Declination. $P_{J,H,K_s}$ $\pm$ $e_{P_{J,H,K_s}}$ and $\theta_{J,H,K_s}$ $\pm$ $e_{\theta_{J,H,K_s}}$ represent the degree and position angle of polarization with its corresponding uncertainties in the $J$, $H$ and $K_s$ bands, respectively.}
\label{Tab:NIR_pol}

\resizebox{1\textwidth}{!}{%
\begin{tabular}{ccccccccc}
\hline
\hline
ID &  R.A & Dec. & $P_{J}$$\pm$ $e_{P_J}$ & $\theta_{J}$ $\pm$ $e_{\theta_J}$  &  $P_{H}$ $\pm$
$e_{P_H}$ & $\theta_{H} $$\pm$ $e_{\theta_H}$ & $P_{K_s}$$\pm$  $e_{P_{K_s}}$ & $\theta_{K_s} $
$\pm$ $e_{\theta_{k_s}}$  \\
&(hh:mm:ss)&(dd:mm:ss)&(per cent)&$(^{\circ})$&(per cent)&$(^{\circ})$&(per cent)&$(^{\circ})$\\
\hline

01$^{\text{m}}$ &  06\ 08\ 37.76 &  21\ 30\ 36.6 &   2.21$\pm$   0.73 & 125$\pm$   10&   1.80$\pm$   0.46&  131$\pm$    7&     1.42$\pm$   0.44&  100$\pm$   22 \\ 
02 &  06\ 08\ 37.67 &  21\ 30\ 41.1 &   4.00$\pm$   1.13 & 136$\pm$    7&   3.35$\pm$   0.53&  142$\pm$    4&     2.16$\pm$   0.54&  162$\pm$    7 \\
03$^{\text{m}}$ &  06\ 08\ 38.28 &  21\ 30\ 39.4 &   1.26$\pm$   0.21 & 145$\pm$    5&   0.66$\pm$   0.15&  140$\pm$    6&  \---   $\pm$  \--- &\---$\pm$ \---  \\
04 &  06\ 08\ 37.40 &  21\ 31\ 08.4 &   7.30$\pm$   1.46 & 139$\pm$    5&   2.49$\pm$   0.55&  132$\pm$    6&     1.69$\pm$   0.47&  147$\pm$    7 \\
05 &  06\ 08\ 38.08 &  21\ 31\ 12.9 &   0.79$\pm$   0.21 & 168$\pm$    8&  \--- $\pm$  \--- &\---$\pm$  \---&  \---   $\pm$  \--- &\---$\pm$ \--- \\
06 &  06\ 08\ 39.82 &  21\ 30\ 53.1 &   1.78$\pm$   0.58 &  52$\pm$   14&   1.70$\pm$   0.55&  159$\pm$   10&     1.49$\pm$   0.51&    9$\pm$   3 \\
07 &  06\ 08\ 39.53 &  21\ 31\ 21.4 &   5.62$\pm$   1.04 & 135$\pm$    9&   2.24$\pm$   0.70&   60$\pm$    9&     4.67$\pm$   0.75&   33$\pm$    4 \\
08 &  06\ 08\ 41.21 &  21\ 31\ 18.6 &   7.80$\pm$   1.34 & 155$\pm$    4&   9.69$\pm$   0.51&  137$\pm$    1&    10.31$\pm$   0.39&  137$\pm$    1 \\
09 &  06\ 08\ 41.88 &  21\ 31\ 12.9 &   1.45$\pm$   0.45 &  75$\pm$   21&   1.32$\pm$   0.31&  129$\pm$   10&     3.17$\pm$   0.39&  154$\pm$    3 \\
10 &  06\ 08\ 39.35 &  21\ 31\ 47.1 &   8.45$\pm$   1.62 & 140$\pm$    5&   4.09$\pm$   0.72&  131$\pm$    4&     2.38$\pm$   0.68&  135$\pm$    7 \\
11 &  06\ 08\ 38.06 &  21\ 32\ 21.6 &   2.50$\pm$   0.74 & 141$\pm$    8&   1.95$\pm$   0.51&  133$\pm$    7&     1.12$\pm$   0.34&  112$\pm$   28 \\
12 &  06\ 08\ 40.66 &  21\ 31\ 55.3 &   2.91$\pm$   0.90 & 146$\pm$    9&   1.66$\pm$   0.49&  126$\pm$    9&     1.65$\pm$   0.50&  165$\pm$   11 \\
13 &  06\ 08\ 43.04 &  21\ 31\ 46.8 &   2.34$\pm$   0.75 & 145$\pm$   11&   1.50$\pm$   0.49&  126$\pm$    8&     1.47$\pm$   0.44&  157$\pm$   19 \\
14 &  06\ 08\ 43.58 &  21\ 31\ 51.1 &   1.44$\pm$   0.45 & 166$\pm$    7&   4.53$\pm$   0.80&  169$\pm$    5&     3.05$\pm$   0.55&  149$\pm$    5 \\
15 &  06\ 08\ 46.12 &  21\ 31\ 45.0 &   4.21$\pm$   1.03 &  75$\pm$   12&   2.46$\pm$   0.74&  102$\pm$   10&     0.74$\pm$   0.18&   90$\pm$   15 \\
16 &  06\ 08\ 49.86 &  21\ 31\ 34.3 &   3.04$\pm$   0.70 & 179$\pm$    6&   2.89$\pm$   0.52&  178$\pm$    5&     3.94$\pm$   0.71&    6$\pm$    2 \\
17$^{\text{m}}$ &  06\ 08\ 47.60 &  21\ 30\ 58.3 &   5.12$\pm$   0.81 & 147$\pm$    4&   3.49$\pm$   0.52&  149$\pm$    4&     2.57$\pm$   0.62&  144$\pm$    6 \\
18 &  06\ 08\ 47.42 &  21\ 30\ 54.8 &   8.01$\pm$   1.92 & 148$\pm$    6&   4.38$\pm$   0.76&  150$\pm$    4&     3.17$\pm$   0.74&  157$\pm$    6 \\
19 &  06\ 08\ 47.05 &  21\ 30\ 04.7 &   2.33$\pm$   0.36 & 141$\pm$    4&   1.40$\pm$   0.23&  137$\pm$    4&     0.73$\pm$   0.20&  158$\pm$   10 \\
20 &  06\ 08\ 44.65 &  21\ 29\ 32.7 &   1.46$\pm$   0.13 &   6$\pm$    2&   0.78$\pm$   0.09&    5$\pm$    2&     0.85$\pm$   0.12&   14$\pm$    4 \\
21$^{\text{m}}$ &  06\ 08\ 44.08 &  21\ 30\ 30.5 &   3.69$\pm$   0.71 & 159$\pm$    5&   2.19$\pm$   0.42&  157$\pm$    5&     1.14$\pm$   0.38&  169$\pm$   11 \\
22 &  06\ 08\ 39.59 &  21\ 30\ 00.2 &   1.44$\pm$   0.35 & 150$\pm$   13&   0.84$\pm$   0.22&  122$\pm$   12&     0.44$\pm$   0.13&   48$\pm$   10 \\
23 &  06\ 08\ 34.78 &  21\ 31\ 58.3 &   2.52$\pm$   0.67 & 139$\pm$    7&   2.32$\pm$   0.35&  135$\pm$    4&     1.48$\pm$   0.39&  140$\pm$    7 \\
24 &  06\ 08\ 33.49 &  21\ 32\ 01.8 &   3.33$\pm$   0.83 & 120$\pm$    9&   1.91$\pm$   0.53&  107$\pm$    9&     3.03$\pm$   0.74&  110$\pm$    6 \\
25 &  06\ 08\ 31.28 &  21\ 32\ 02.1 &   1.59$\pm$   0.43 & 124$\pm$   11&   0.88$\pm$   0.25&  124$\pm$   11&     0.78$\pm$   0.18&  136$\pm$   15 \\
26 &  06\ 08\ 30.08 &  21\ 32\ 28.8 &   1.73$\pm$   0.28 & 135$\pm$    4&   0.98$\pm$   0.21&  135$\pm$    6&     0.63$\pm$   0.18&  123$\pm$   11 \\
27 &  06\ 08\ 30.99 &  21\ 31\ 39.6 &   1.37$\pm$   0.23 & 170$\pm$    9&   0.99$\pm$   0.24&  172$\pm$   10&     0.80$\pm$   0.25&  144$\pm$   16 \\
28 &  06\ 08\ 32.58 &  21\ 30\ 17.9 &   1.12$\pm$   0.32 & 127$\pm$   15&   1.09$\pm$   0.30&  121$\pm$    9&     0.93$\pm$   0.26&  132$\pm$   12 \\ 
29$^{\text{m}}$ &  06\ 08\ 26.72 &  21\ 29\ 59.1 &  \--- $\pm$  \---  &\---$\pm$ \---&   1.03$\pm$   0.23&  110$\pm$   14&     2.63$\pm$   0.77&  146$\pm$    8 \\
30 &  06\ 08\ 28.99 &  21\ 28\ 30.1 &   1.28$\pm$   0.18 & 125$\pm$    4&   0.83$\pm$   0.12&  121$\pm$    4&     0.50$\pm$   0.16&  108$\pm$    8 \\
31$^{\text{m}}$ &  06\ 08\ 27.97 &  21\ 28\ 23.6 &   1.28$\pm$   0.36 & 146$\pm$   12&   0.90$\pm$   0.22&  121$\pm$   14&     0.56$\pm$   0.15&  119$\pm$   14 \\
32 &  06\ 08\ 35.55 &  21\ 28\ 22.2 &   1.44$\pm$   0.40 & 167$\pm$    9&   0.81$\pm$   0.25&  172$\pm$   12&     0.52$\pm$   0.16&   36$\pm$   10 \\
33 &  06\ 08\ 32.26 &  21\ 28\ 09.2 &  \--- $\pm$  \---  &\---$\pm$ \---&   0.62$\pm$   0.22&  139$\pm$   36&     1.92$\pm$   0.60&   20$\pm$   7 \\
34 &  06\ 08\ 37.96 &  21\ 28\ 01.0 &   5.28$\pm$   1.22 & 143$\pm$    6&   3.11$\pm$   0.56&  139$\pm$    5&     1.27$\pm$   0.29&  154$\pm$   12 \\
35 &  06\ 08\ 40.18 &  21\ 28\ 58.0 &   2.68$\pm$   0.71 & 147$\pm$    7&   1.74$\pm$   0.43&  146$\pm$    6&     1.14$\pm$   0.32&  151$\pm$   12 \\
36 &  06\ 08\ 40.83 &  21\ 29\ 12.5 &   0.95$\pm$   0.27 & 175$\pm$    8&   0.46$\pm$   0.14&  175$\pm$   13&     0.81$\pm$   0.26&    6$\pm$    8 \\
37 &  06\ 08\ 46.61 &  21\ 27\ 44.6 &   1.85$\pm$   0.64 &   3$\pm$    9&   0.72$\pm$   0.37&    2$\pm$   11&     2.78$\pm$   0.66&    1$\pm$    6 \\
38 &  06\ 08\ 39.38 &  21\ 27\ 49.1 &  \--- $\pm$  \---  &\---$\pm$ \---&  \--- $\pm$  \--- &\---$\pm$  \---&     3.09$\pm$   1.00&  173$\pm$    8 \\ 
39 &  06\ 08\ 50.34 &  21\ 28\ 11.6 &   3.27$\pm$   0.37 & 150$\pm$    3&   1.88$\pm$   0.17&  149$\pm$    2&     1.45$\pm$   0.19&  170$\pm$    3 \\
40$^{\text{m}}$ &  06\ 08\ 50.66 &  21\ 28\ 22.6 &   2.37$\pm$   0.55 & 147$\pm$    6&   1.40$\pm$   0.39&  141$\pm$    7&     1.32$\pm$   0.41&  169$\pm$   10 \\
41 &  06\ 08\ 52.52 &  21\ 28\ 44.6 &   0.73$\pm$   0.33 &  16$\pm$    8&   0.67$\pm$   0.20&   74$\pm$   15&     1.13$\pm$   0.35&   14$\pm$   10 \\
42 &  06\ 08\ 51.95 &  21\ 28\ 13.6 &   1.76$\pm$   0.49 & 172$\pm$    8&   1.04$\pm$   0.34&  172$\pm$    8&     0.87$\pm$   0.25&  179$\pm$   13 \\
43 &  06\ 08\ 52.74 &  21\ 27\ 36.6 &   0.50$\pm$   0.15 &  42$\pm$   10&   0.44$\pm$   0.13&   56$\pm$   16&     1.34$\pm$   0.33&   20$\pm$    6 \\
44 &  06\ 08\ 54.49 &  21\ 28\ 17.1 &   5.97$\pm$   0.87 & 149$\pm$    4&   3.94$\pm$   0.38&  150$\pm$    2&     2.25$\pm$   0.39&  167$\pm$    4 \\
45 &  06\ 08\ 54.80 &  21\ 30\ 17.6 &   0.94$\pm$   0.30 & 173$\pm$   23&   0.74$\pm$   0.17&   11$\pm$   10&     0.59$\pm$   0.15&  136$\pm$   15 \\
46 &  06\ 08\ 56.91 &  21\ 30\ 57.1 &   0.55$\pm$   0.15 &  62$\pm$   12&   0.70$\pm$   0.15&   59$\pm$    6&     1.19$\pm$   0.20&   28$\pm$    4 \\
47 &  06\ 08\ 54.94 &  21\ 32\ 30.1 &   0.72$\pm$   0.20 & 175$\pm$   16&   0.45$\pm$   0.15&  171$\pm$   18&     1.44$\pm$   0.45&  175$\pm$   10 \\
48 &  06\ 08\ 54.91 &  21\ 33\ 14.6 &   1.07$\pm$   0.33 & 176$\pm$    9&   0.81$\pm$   0.20&  177$\pm$    8&     0.87$\pm$   0.24&   10$\pm$   10 \\
49$^{\text{m}}$ &  06\ 08\ 49.50 &  21\ 33\ 54.2 &   2.35$\pm$   0.54 & 159$\pm$    6&   1.28$\pm$   0.35&  158$\pm$    7&     0.63$\pm$   0.14&  148$\pm$   16 \\
50 &  06\ 08\ 47.09 &  21\ 34\ 37.3 &   2.85$\pm$   0.35 & 174$\pm$    3&   1.42$\pm$   0.29&  178$\pm$    5&     0.91$\pm$   0.30&  172$\pm$   11 \\
51 &  06\ 08\ 42.10 &  21\ 34\ 33.8 &   2.33$\pm$   0.37 & 148$\pm$    4&   1.34$\pm$   0.29&  146$\pm$    6&     0.79$\pm$   0.19&  147$\pm$   12 \\
52$^{\text{m}}$ &  06\ 08\ 42.96 &  21\ 33\ 52.8 &   2.36$\pm$   0.66 & 149$\pm$    8&   1.08$\pm$   0.31&  150$\pm$    9&     0.64$\pm$   0.20&  137$\pm$   21 \\
53 &  06\ 08\ 42.75 &  21\ 33\ 14.3 &   1.17$\pm$   0.33 &   3$\pm$    5&   0.54$\pm$   0.15&    6$\pm$   10&     0.49$\pm$   0.15&    3$\pm$   10 \\
54 &  06\ 08\ 42.39 &  21\ 32\ 56.3 &   0.96$\pm$   0.31 & 162$\pm$   10&   0.65$\pm$   0.20&  172$\pm$   10&     0.51$\pm$   0.16&    1$\pm$   10 \\
55 &  06\ 08\ 45.01 &  21\ 33\ 12.8 &   1.86$\pm$   0.55 & 151$\pm$   12&   0.53$\pm$   0.16&  168$\pm$   19&     0.55$\pm$   0.13&    9$\pm$   15 \\
56 &  06\ 08\ 38.73 &  21\ 33\ 26.8 &   1.59$\pm$   0.22 & 171$\pm$    4&   0.94$\pm$   0.17&  173$\pm$    5&     0.62$\pm$   0.20&    6$\pm$   10 \\
57 &  06\ 08\ 38.73 &  21\ 33\ 12.8 &   4.13$\pm$   0.39 & 146$\pm$    2&   2.51$\pm$   0.20&  145$\pm$    2&     1.59$\pm$   0.24&  147$\pm$    4 \\
58 &  06\ 08\ 39.63 &  21\ 33\ 13.8 &   1.80$\pm$   0.43 & 139$\pm$   10&   1.85$\pm$   0.44&  141$\pm$    6&     0.46$\pm$   0.14&  158$\pm$   21 \\
59 &  06\ 08\ 35.36 &  21\ 34\ 35.3 &   1.64$\pm$   0.50 & 137$\pm$    9&   1.38$\pm$   0.43&  132$\pm$    8&     0.45$\pm$   0.15&  159$\pm$   22 \\
60 &  06\ 08\ 35.11 &  21\ 34\ 12.3 &   1.42$\pm$   0.28 & 169$\pm$    8&   1.97$\pm$   0.46&  135$\pm$    9&     2.34$\pm$   0.69&  146$\pm$    8 \\
61 &  06\ 08\ 35.26 &  21\ 33\ 47.3 &   0.52$\pm$   0.17 & 166$\pm$   20&   0.84$\pm$   0.25&  151$\pm$   16&     0.85$\pm$   0.26&   51$\pm$   12 \\ 
62 &  06\ 08\ 30.38 &  21\ 32\ 50.8 &   3.07$\pm$   0.42 & 134$\pm$    4&   2.01$\pm$   0.24&  132$\pm$    3&     1.51$\pm$   0.29&  129$\pm$    5 \\
63 &  06\ 08\ 28.99 &  21\ 33\ 13.8 &   1.47$\pm$   0.45 & 176$\pm$   13&   0.97$\pm$   0.26&    3$\pm$    7&     1.37$\pm$   0.37&   11$\pm$   10 \\
64 &  06\ 08\ 27.52 &  21\ 32\ 26.3 &   2.59$\pm$   0.66 & 157$\pm$    9&   1.56$\pm$   0.50&  150$\pm$    8&     1.87$\pm$   0.61&  127$\pm$    8 \\
\hline
\end{tabular}
}
\setlength{\leftskip}{-25pt}\\
\setlength{\leftskip}{30pt}
\small{Notes: $^{\text{m}}$ denotes Member stars identified from \href{https://academic.oup.com/mnras/article/541/2/1557/8173845}{Paper I}}
\end{table*}


\begin{table*}[t]\hspace{-20cm}
\setlength{\tabcolsep}{12pt} 
\centering
\caption{Observed $B, V, R_c$ and $I_c$ polarization values for 31 stars in AFGL 6366S. The symbols have the same meaning as in Table~\ref{Tab:NIR_pol}, but for the $B$, $V$, $R_c$ and $I_c$ photometric bands.}
\label{Tab:Opt_pol}
\resizebox{\textwidth}{0.3\textheight}{%
\begin{tabular}{ccccccccc} 
\hline\hline
ID & $P_{B}$ $\pm$ $e_{P_{\text{B}}}$  & $\theta_{B}$ $\pm$ $e_{\theta_{B}}$  &  $P_{V}$ $\pm$
$e_{P_{\text{V}}}$ & $\theta_{V} $$\pm$ $e_{\theta_{V}}$ & $P_{R}$ $\pm$  $e_{P_{\text{R}}}$ & $\theta_{R} $
$\pm$ $e_{\theta_{R}}$ & $P_{I}$$\pm$  $e_{P_{\text{I}}}$ & $\theta_{I} $ $\pm$ $e_{\theta_{I}}$\\
&(per cent)&$(^{\circ})$&(per cent)&$(^{\circ})$&(per cent)&$(^{\circ})$&(per cent)&$(^{\circ})$\\
\hline                                                         \\ 
\hline
05     & 1.90 $\pm$ 0.03                       & 175 $\pm$ 2                              & 2.04 $\pm$ 0.01                      & 171 $\pm$ 2                       & 9.97 $\pm$ 1.12                       & 5 $\pm$ 2                                & 5.02 $\pm$ 0.63                 & 34 $\pm$ 13                                \\
11     & -- $\pm$ --                            & -- $\pm$ --                               & 5.13 $\pm$ 0.53                      & 150 $\pm$ 2                       & 5.19 $\pm$ 0.34                       & 135 $\pm$ 2                              & 8.52 $\pm$ 0.71                 & 105$\pm$ 10                                \\
17$^{\text{m}}$ & 1.69 $\pm$ 0.03                       & 93 $\pm$ 2                               & 0.31 $\pm$ 0.17                      & 101 $\pm$ 5                       & 0.23 $\pm$ 0.10                       & 123 $\pm$ 7                              & 0.44 $\pm$ 0.00                 & 153$\pm$ 2                                 \\
18     & 2.78 $\pm$ 0.10                       & 15 $\pm$ 2                               & 6.24 $\pm$ 0.43                      & 2 $\pm$ 2                         & 5.53 $\pm$ 0.02                       & 4$\pm$ 2                                 & 5.97 $\pm$ 0.23                 & 5$\pm$ 2                                   \\
21$^{\text{m}}$ & 1.82 $\pm$ 0.51                       & 171 $\pm$ 4                              & 2.43 $\pm$ 0.22                      & 179 $\pm$ 2                       & 2.54 $\pm$ 0.10                       & 176 $\pm$ 2                              & 2.13 $\pm$ 0.17                 & 176$\pm$ 2                                 \\
22     & 8.00 $\pm$ 1.52                       & 36 $\pm$ 10                              & 3.09 $\pm$ 0.36                      & 17 $\pm$ 3                        & 1.31 $\pm$ 0.37                       & 51$\pm$ 8                                & 1.32 $\pm$ 0.43                 & 16$\pm$ 9                                  \\
23     & 8.25 $\pm$ 2.84                       & 137 $\pm$ 12                             & 6.35 $\pm$ 0.10                      & 148 $\pm$ 2                       & 6.12 $\pm$ 0.12                       & 144 $\pm$ 2                              & 5.06 $\pm$ 0.05                 & 147$\pm$ 2                                 \\
24     & 2.31 $\pm$ 0.02                       & 2 $\pm$ 2                                & 2.88 $\pm$ 0.18                      & 2 $\pm$ 2                         & 1.91 $\pm$ 0.38                       & 168 $\pm$ 2                              & 7.53 $\pm$ 0.11                 & 16$\pm$ 2                                  \\
29$^{\text{m}}$ & -- $\pm$ --                           & -- $\pm$ --                               & -- $\pm$ --                          & -- $\pm$ --                        & 2.20 $\pm$ 0.74                       & 151 $\pm$ 9                              & 2.84 $\pm$ 0.62                 & 127$\pm$ 8                                 \\
30     & 2.34 $\pm$ 0.20                       & 160 $\pm$ 2                              & 3.60 $\pm$ 0.32                      & 147 $\pm$ 2                       & 3.07 $\pm$ 0.03                       & 144 $\pm$ 2                              & 2.84 $\pm$ 0.11                 & 144$\pm$ 2                                 \\
31$^{\text{m}}$ & 3.31 $\pm$ 0.02                       & 179 $\pm$ 2                              & 3.57 $\pm$ 0.27                      & 4 $\pm$ 2                         & 3.19 $\pm$ 0.13                       & 173 $\pm$ 2                              & 3.14 $\pm$ 0.12                 & 177$\pm$ 2                                 \\
32     & -- $\pm$ --                           & -- $\pm$ --                               & 3.72 $\pm$ 0.12                      & 9 $\pm$ 2                         & 3.02 $\pm$ 0.62                       & 157 $\pm$ 5                              & 5.19 $\pm$ 0.08                 & 130$\pm$ 2                                 \\
33     & 5.15 $\pm$ 1.51                       & 66 $\pm$ 8                               & 0.22 $\pm$ 0.28                      & 130 $\pm$ 7                       & 0.69 $\pm$ 0.22                       & 120 $\pm$ 5                              & 0.35 $\pm$ 0.27                 & 14$\pm$ 6                                  \\
34     & -- $\pm$ --                           & -- $\pm$ --                               & -- $\pm$ --                          & -- $\pm$ --                         & 4.97 $\pm$ 0.94                       & 159 $\pm$ 5                              & 3.01 $\pm$ 0.51                 & 7$\pm$ 10                                  \\
35     & -- $\pm$ --                           & -- $\pm$--                               & -- $\pm$ --                          & -- $\pm$ --                         & 2.64 $\pm$ 0.01                       & 129 $\pm$ 2                              & 2.31 $\pm$ 0.16                 & 141$\pm$ 2                                 \\
36     & -- $\pm$ --                           & -- $\pm$ --                               & -- $\pm$ --                          & --$\pm$--                         & 3.90 $\pm$ 1.06                       & 132 $\pm$ 7                              & 1.44 $\pm$ 0.00                 & 7$\pm$ 2                                   \\
37     & -- $\pm$ --                           & -- $\pm$--                               & -- $\pm$ --                          & -- $\pm$ --                         & 4.29 $\pm$ 0.49                       & 34 $\pm$ 3                               & 0.61 $\pm$ 0.51                 & 52 $\pm$ 9                                 \\
38     & 3.77 $\pm$ 0.62                       & 2 $\pm$ 4                                & 1.95 $\pm$ 0.21                      & 178 $\pm$ 3                       & 1.97 $\pm$ 0.07                       & 168 $\pm$ 2                              & 2.38 $\pm$ 0.45                 & 169$\pm$ 5                                 \\
39     & -- $\pm$ --                           & -- $\pm$--                               & -- $\pm$ --                          & -- $\pm$ --                         & 2.09 $\pm$ 0.92                       & 2 $\pm$ 9                                & 11.12 $\pm$ 1.65                & 138$\pm$ 4                                 \\
42     & -- $\pm$ --                           & -- $\pm$ --                               & 3.78 $\pm$ 0.02                      & 153 $\pm$ 2                       & 2.87 $\pm$ 0.15                       & 153 $\pm$ 2                              & 3.57$\pm$ 1.52                  & 173$\pm$ 12                                \\
43     & 3.06 $\pm$ 0.26                       & 2 $\pm$ 2                                & 1.96 $\pm$ 0.02                      & 1 $\pm$ 2                         & 2.33 $\pm$ 0.01                       & 177 $\pm$ 2                              & 2.45 $\pm$ 0.13                 & 179$\pm$ 2                                 \\
45     & 3.34 $\pm$ 0.98                       & 9 $\pm$ 8                                & 4.19 $\pm$ 0.38                      & 179 $\pm$ 2                       & 3.83 $\pm$ 0.14                       & 176 $\pm$ 2                              & 5.69$\pm$ 0.31                  & 133$\pm$ 8                                 \\
49$^{\text{m}}$ & 5.51 $\pm$ 1.41                       & 133 $\pm$ 7                              & 5.81 $\pm$ 0.01                      & 154 $\pm$ 2                       & 5.34 $\pm$ 0.16                       & 136$\pm$ 2                               & 11.22 $\pm$ 2.84                & 113$\pm$ 7                                 \\
51     & 3.05 $\pm$ 0.30                       & 175 $\pm$ 2                              & 4.84 $\pm$ 0.18                      & 172 $\pm$ 2                       & 4.46 $\pm$ 0.29                       & 162 $\pm$ 2                              & 8.86 $\pm$ 1.41                 & 174$\pm$ 4                                 \\
54     & 4.25 $\pm$ 0.88                       & 11 $\pm$ 3                               & 2.14 $\pm$ 0.12                      & 2 $\pm$ 2                         & 2.60 $\pm$ 0.64                       & 175$\pm$ 7                               & 2.70 $\pm$ 0.87                 & 90$\pm$ 9                                  \\
55     & 5.37 $\pm$ 1.40                       & 37 $\pm$ 10                              & 2.92 $\pm$ 0.09                      & 1 $\pm$ 2                         & 2.57 $\pm$ 0.10                       & 2 $\pm$ 2                                & 1.51 $\pm$ 0.11                 & 174$\pm$ 2                                 \\
56     & 0.47 $\pm$ 0.03                       & 85 $\pm$ 2                               & 0.36 $\pm$ 0.16                      & 44 $\pm$ 3                        & 0.17 $\pm$ 0.01                       & 34$\pm$ 2                                & 0.16 $\pm$ 0.03                 & 12$\pm$ 5                                  \\
57     & 2.41 $\pm$ 0.01                       & 172 $\pm$ 2                              & 3.19 $\pm$ 0.34                      & 2 $\pm$ 3                         & 2.29 $\pm$ 0.03                       & 170$\pm$ 2                               & 1.85 $\pm$ 0.27                 & 169$\pm$ 4                                 \\
58     & 1.12 $\pm$ 0.41                       & 19 $\pm$ 8                               & 3.37 $\pm$ 0.30                      & 166 $\pm$ 5                       & 2.08 $\pm$ 0.27                       & 167$\pm$ 3                               & 2.58 $\pm$ 0.17                 & 164$\pm$ 2                                 \\
62     & 5.54 $\pm$ 0.31                       & 28 $\pm$ 2                               & 7.98 $\pm$ 0.93                      & 170 $\pm$ 3                       & 4.78 $\pm$ 0.65                       & 169$\pm$ 3                               & 4.31 $\pm$ 0.17                 & 161$\pm$ 2                                 \\
64     & 3.26 $\pm$ 0.08                       & 169 $\pm$ 2                              & 1.87 $\pm$ 0.18                      & 174 $\pm$ 2                       & 0.98 $\pm$ 0.73                       & 156$\pm$ 30                              & 2.09 $\pm$ 0.30                 & 5$\pm$ 4                                   \\
\hline
\end{tabular}
}
\setlength{\leftskip}{-25pt}\\
\end{table*}

Table \ref{Tab:NIR_pol} lists the NIR ($JHK_s$) polarimetric measurements for the 64 stars towards AFGL 6366S. 
The complementary Optical ($BVR_cI_c$) polarimetric measurements are provided in Table \ref{Tab:Opt_pol} for 31 out of these 64 stars. It should be noted that polarization angles differing by 180$^\circ$ represent the same polarization orientation \citep[e.g.,][]{singh_2020,Biswas_2024}. Therefore, in all the Tables that follow, $\theta > 180^\circ$ are reported as $(180 - \theta)^\circ$. 

Among the 64 observed sources, nine (IDs \#01, \#03, \#17, \#21, \#29, \#31, \#40, \#49, and \#52) sources were identified as cluster members from the membership analysis of AFGL 6366S presented in \href{https://academic.oup.com/mnras/article/541/2/1557/8173845}{Paper I}. These are marked with a superscript "m" in Tables \ref{Tab:NIR_pol}, \ref{Tab:Opt_pol} and \ref{tab:ser_law}.

\begin{figure}\hspace{0cm}
	\includegraphics[trim=0cm 0cm 0cm 0cm, clip,width=\columnwidth]{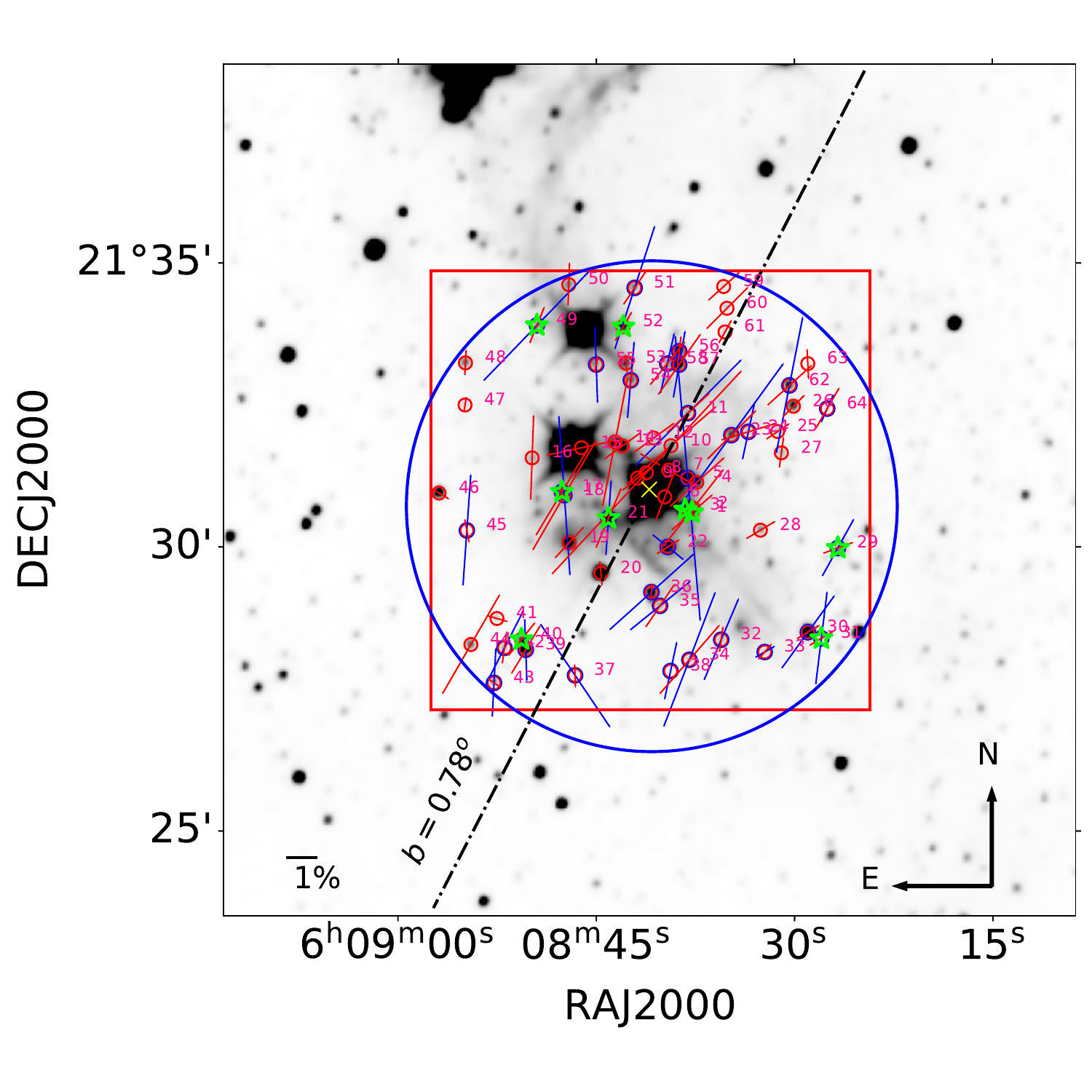}
	\caption{Polarization map of AFGL 6366S showing the NIR (red square) and Optical (large blue circle) polarimetric observation fields superimposed on a 15$'$ × 15$'$ WISE 3.4 $\mu m$ image. The sources observed in NIR and Optical are marked by small red and blue circles, respectively. The observed star IDs according to Table \ref{Tab:NIR_pol} and/or Table \ref{Tab:Opt_pol}, are annotated. The green star marks represents the cluster members identified from  \href{https://academic.oup.com/mnras/article/541/2/1557/8173845}{Paper I}. The red and/or blue polarization vectors on each star represent their $H$ band and/or $R_c$ band polarization, respectively. The length of the polarization vectors is proportional to degree of polarization (in per cent). The inclination of the polarization vectors with respect to the north-south direction indicates the position angle of polarization. A line of magnitude of 1 per cent polarization value is drawn for reference. The dashed black line indicates the orientation of the projection of the Galactic plane at b = $0.78^{\circ}$. The yellow cross indicates the centre of the cluster AFGL 6366S at R.A. (J2000) = $06^h08^m41^s$ and Dec. (J2000) = $+21^{\circ}31'01''$. }
	\label{Fig: pol}
\end{figure}

Fig. \ref{Fig: pol} presents the sky-projected polarization vectors of the observed stars in the $H$ and the $R_c$ photometric bands, in red and blue colours, respectively (the polarization vectors from the remaining photometric bands are not shown to avoid clutter, but they exhibit a broadly similar pattern). The dashed black line signifies the orientation of the Galactic plane (GP) projection at b = $0.78^{\circ}$. The position angle of the GP ($\theta_{\text{GP}}$) with respect to the North-South direction is $151.04^\circ$. 

It may be observed in Fig. \ref{Fig: pol} that, for stars with polarization measurements in both $H$ and $R_c$ bands, the corresponding vectors exhibit slight offsets in orientation despite representing the same source. Such small differences are generally expected and likely arise from a combination of wavelength-dependent effects and/or instrumental characteristics.

An overview of the polarization properties across all the observed photometric bands is provided in Table \ref{tab:pol_summary}. It reports the ranges of the parameters $P$ and $\theta$ for the entire observed sample and its subset of the cluster members. It also lists the inverse-variance-weighted averages of $P$ and $\theta$ for the entire sample ($\langle P \rangle$ and $\langle \theta \rangle$) and the member population ($\langle P \rangle_{\text{members}}$ and $\langle \theta \rangle_{\text{members}}$) 

\begin{figure*}
\begin{subfigure}{\columnwidth}\hspace{0.5cm}
\includegraphics[width=0.9\columnwidth]{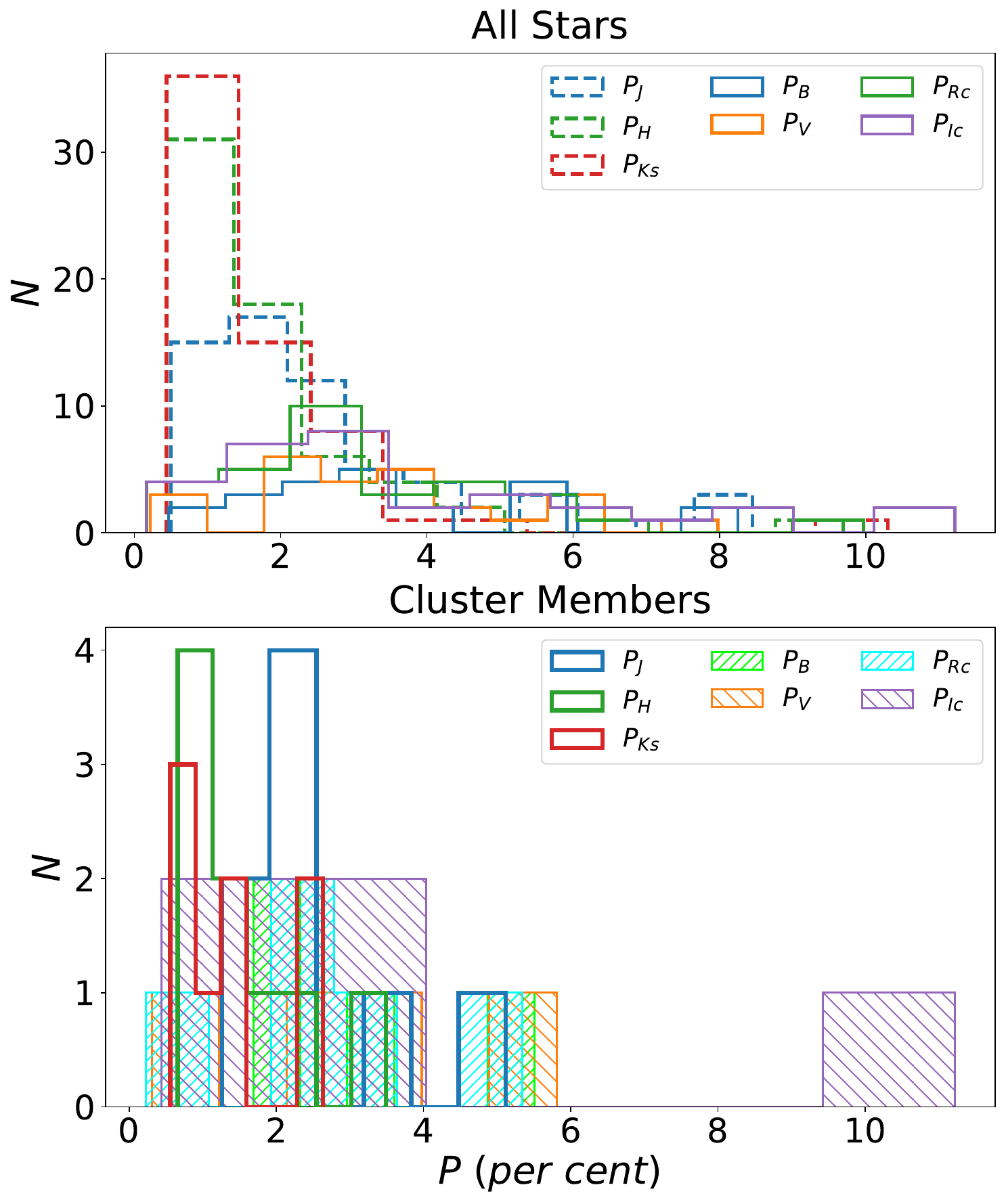}
\caption{}
\label{Fig:Opt_NIR_pol_hist}
\end{subfigure}
\hfil
\begin{subfigure}{\columnwidth}\hspace{0.5cm}
\includegraphics[width=0.9\columnwidth]{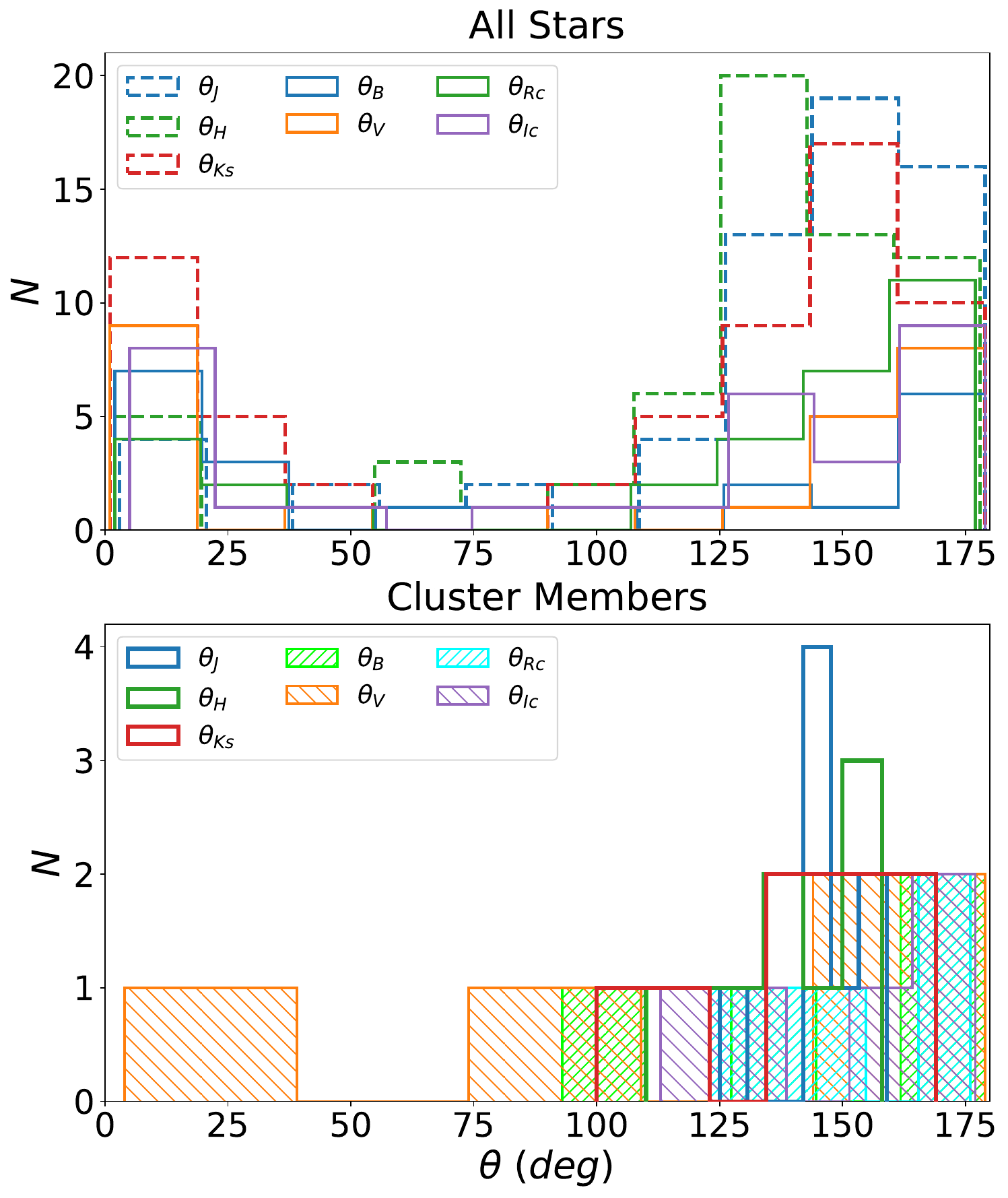}
\caption{}
\label{Fig:Opt_NIR_theta_hist}
\end{subfigure}
\caption{(a) Histograms of polarization degree (per cent) for all stars (in upper panel) and member stars (in lower panel) in $JHK_sBVR_cI_c$ bands. (b) Histograms of polarization angle (degrees) for all stars (in upper panel) and member stars (in lower panel) in $JHK_sBVR_cI_c$ bands.}
\label{Fig:pol_pa_hist}
\end{figure*}

Fig. \ref{Fig:pol_pa_hist} illustrates the distributions of $P$ and $\theta$ across the NIR and Optical bands for all the observed stars as well as the member population.

\begin{table*}
\setlength{\tabcolsep}{3pt}
\centering
\caption{Summary of polarization properties for all observed stars and confirmed members of AFGL 6366S. Columns 2–3 list the ranges of $P$ and $\theta$ for all stars across different photometric filters; Columns 4–5 give the corresponding ranges for member stars. Columns 6–7 present the error-weighted mean values of $P$ and $\theta$ for all stars, while Columns 8–9 list those for the member subset.}
\label{tab:pol_summary}
\renewcommand{\arraystretch}{1.3}
\begin{tabular}{ccccccccc}
\hline
Band & $P$ Range & $\theta$ Range & $P$ Range (Members) & $\theta$ Range (Members) & $\langle P \rangle$ & $\langle \theta \rangle$ & $\langle P \rangle_{\text{members}}$  & $\langle \theta \rangle_{\text{members}}$ \\
& (per cent) & $(^{\circ})$ & (per cent) & $(^{\circ})$ & (per cent) & $(^{\circ})$ & (per cent) & $(^{\circ})$ \\
\hline
$J$     & 0.50--8.45 & 3--179 & 1.26--5.12 & 125--159 & 1.36 $\pm$ 0.29 & 155.00 $\pm$ 4.65 & 1.76 $\pm$ 0.51 & 149.41 $\pm$ 3.17 \\
$H$     & 0.44--9.69 & 2--178 & 0.66--3.49 & 110--158 & 1.03 $\pm$ 0.20 & 145.26 $\pm$ 6.42 & 1.07 $\pm$ 0.31 & 146.47 $\pm$ 4.42 \\
$K_s$   & 0.44--10.3 & 1--179 & 0.56--2.63 & 100--169 & 0.89 $\pm$ 0.21 & 154.92 $\pm$ 12.62 & 0.75 $\pm$ 0.35 & 148.02 $\pm$ 7.38 \\
\textit{B}   & 0.47--8.25 & 2--179 & 1.69--5.51 & 93--179  & 2.34 $\pm$ 0.50 & 0.14 $\pm$ 8.65 & 2.81 $\pm$ 0.78 & 154.24 $\pm$ 56.57 \\
\textit{V}   & 0.22--7.98 & 1--179 & 0.31--5.81 & 4--179   & 3.70 $\pm$ 0.39 & 172.85 $\pm$ 3.78 & 5.78 $\pm$ 1.70 & 171.58 $\pm$ 10.72 \\
\textit{$R_c$} & 0.17--9.97 & 2--177 & 0.23--5.34 & 123--176 & 2.07 $\pm$ 0.41 & 165.62 $\pm$ 5.12 & 2.26 $\pm$ 0.76 & 161.51 $\pm$ 11.40 \\
\textit{I$_c$} & 0.16--11.22 & 5--179 & 0.44--11.22 & 113--177 & 1.07 $\pm$ 0.72 & 166.07 $\pm$ 5.28 & 0.47 $\pm$ 2.29 & 167.44 $\pm$ 8.35 \\
\hline
\end{tabular}
\setlength{\leftskip}{-10pt}\\
\end{table*}

\subsection{Wavelength dependence of polarization}
\label{sec: Ser_law}

The wavelength dependence of polarization is known to be well characterized by the Serkowski law of interstellar polarization \citep{serkowski_1973, serkowski_1975, coyne_1979}. It is expressed as:
\begin{equation}
	P_{\lambda}/    P_{\text{max}}   =\    exp    \left[-\   k    \   ln^{2}    \
	(\lambda_{\text{max}}/\lambda) \right]	
\label{eq:ser_law}
\end{equation}

where, $P_{\lambda}$ is the polarization at wavelength $\lambda$, $P_{\text{max}}$ is the maximum polarization, and $\lambda_{\text{max}}$ denotes the wavelength at which the maximum polarization occurs. The value of $\lambda_{\text{max}}$ depends on the Optical properties as well as the characteristic size distribution of the polarizing dust grains \citep{mcmillan_1978, wilking_1980}. The parameter $P_{\text{max}}$ depends on the column density, the chemical composition, size, shape and alignment efficiency of the dust grains \citep{2008eic..work..511B, 2013AIPC.1543..129L}. The parameter $k$ is a measure of the inverse width of the polarization curve \citep{eswaraiah_2011}.

Polarization measurements from UV to NIR wavelengths are consistent with the predictions of the Serkowski law \citep{2017PASJ...69...25I}. However, the literature reports variations in the adopted values of the parameter $k$. Some studies show that adopting the parameter $k = 1.15 $ provides an adequate representation of the observed polarization in the wavelength range of 0.36 to 1.0 $\mu m$ \citep[e.g.,][]{Medhi_2008, Medhi_2010, eswaraiah_2011, Biswas_2024}. On the other hand, for NIR wavelengths, a better fit may be obtained when $k$ is expressed as a linear function of $\lambda_{\text{max}}$: $k = (-0.10 \pm 0.05) + (1.86 \pm 0.09)\lambda_{\text{max}}$ \citep{1982AJ.....87..695W}. 

Additionally, the wavelength dependence of \textit{JHK$_s$} polarization can also be effectively described by the empirical relation $P \sim \lambda^{-\beta}$ \citep{Nakajima_2007}. Again, a range of different values of  $\beta$, including 2.0 \citep{Nagata_1990}, 1.6 \citep{Martin_1992}, 1.5 to 2 \citep{Martin_whittet_1990}, 1.8 $\pm$ 0.2 \citep{whittet_1992}, and 0.9 \citep{Nakajima_2007} can be found in the literature.

\begin{table}
\setlength{\tabcolsep}{10pt}
\centering
\caption{Serkowski parameters for $BV(RI)c$ polarization toward AFGL 6366S. $P_{\rm max} \pm e_{P_{\rm max}}$ denotes the maximum polarization with its corresponding uncertainty. $\lambda_{\rm max} \pm e_{\lambda_{\rm max}}$ denotes the wavelength of maximum polarization with uncertainty and $\sigma_1$ represents the unit weight error of the Serkowski law fit.}
\label{tab:ser_law}
\renewcommand{\arraystretch}{0.9}
\begin{tabular}{cccc}
\hline
ID & $P_{\text{max}} \pm e_{P_{\text{max}}}$ & $\lambda_{\text{max}} \pm e_{\lambda_{\text{max}}}$ & $\sigma_1$ \\
\hline
5  & 2.05 $\pm$ 0.01   & 0.58 $\pm$ 0.02  & 6.20 \\
18 & 7.11 $\pm$ 0.25   & 1.05 $\pm$ 0.03  & 3.72 \\
21$^{\text{m}}$ & 2.50 $\pm$ 0.11   & 0.60 $\pm$ 0.07  & 0.96 \\
23$^{\text{a}}$ & 6.45 $\pm$ 0.12   & 0.51 $\pm$ 0.01  & 1.51 \\
24 & 10.27 $\pm$ 0.30  & 1.38 $\pm$ 0.02  & 7.93 \\
30 & 3.07 $\pm$ 0.03   & 0.65 $\pm$ 0.04  & 1.70 \\
31$^{\text{m}}$ & 3.52 $\pm$ 0.05   & 0.55 $\pm$ 0.01  & 1.36 \\
38 & 2.30 $\pm$ 0.33   & 0.46 $\pm$ 0.08  & 2.25 \\
43 & 2.57 $\pm$ 0.05   & 0.89 $\pm$ 0.02  & 4.43 \\
45 & 6.32 $\pm$ 1.09   & 1.22 $\pm$ 0.16  & 2.98 \\
49$^{\text{m}}$ & 5.87 $\pm$ 0.09   & 0.50 $\pm$ 0.04  & 1.67 \\
51 & 5.38 $\pm$ 0.49   & 0.81 $\pm$ 0.08  & 2.89 \\
54 & 2.22 $\pm$ 0.18   & 0.52 $\pm$ 0.20  & 1.98 \\
55 & 3.96 $\pm$ 0.42   & 0.34 $\pm$ 0.03  & 2.16 \\
56$^{\text{b}}$ & 1.04 $\pm$ 0.23   & 0.19 $\pm$ 0.02  & 2.18 \\
57 & 2.47 $\pm$ 0.01   & 0.51 $\pm$ 0.01  & 1.55 \\
58$^{\text{a}}$ & 2.95 $\pm$ 0.30   & 0.53 $\pm$ 0.07  & 3.12 \\
62 & 5.74 $\pm$ 0.25   & 0.49 $\pm$ 0.02  & 1.85 \\
64$^{\text{c}}$ & 4.42 $\pm$ 0.72   & 0.26 $\pm$ 0.03  & 4.35 \\
\hline
\end{tabular}
\setlength{\leftskip}{0pt}\\
\setlength{\leftskip}{3pt}
\small{Notes: $P_{\text{max}}$ \& $\lambda_{\text{max}}$ were derived from: $^{\text{a}}$ $VR_cI_c$ polarization data, $^{\text{b}}$ $BR_cI_c$ polarization data, \\ $^{\text{c}}$ $BVI_c$ polarization data.}
\end{table}

\subsubsection{$BVR_cI_c$ polarization}
\label{sec: wave_dep_opt}

\begin{figure}\hspace{0cm}
\includegraphics[width = \columnwidth]{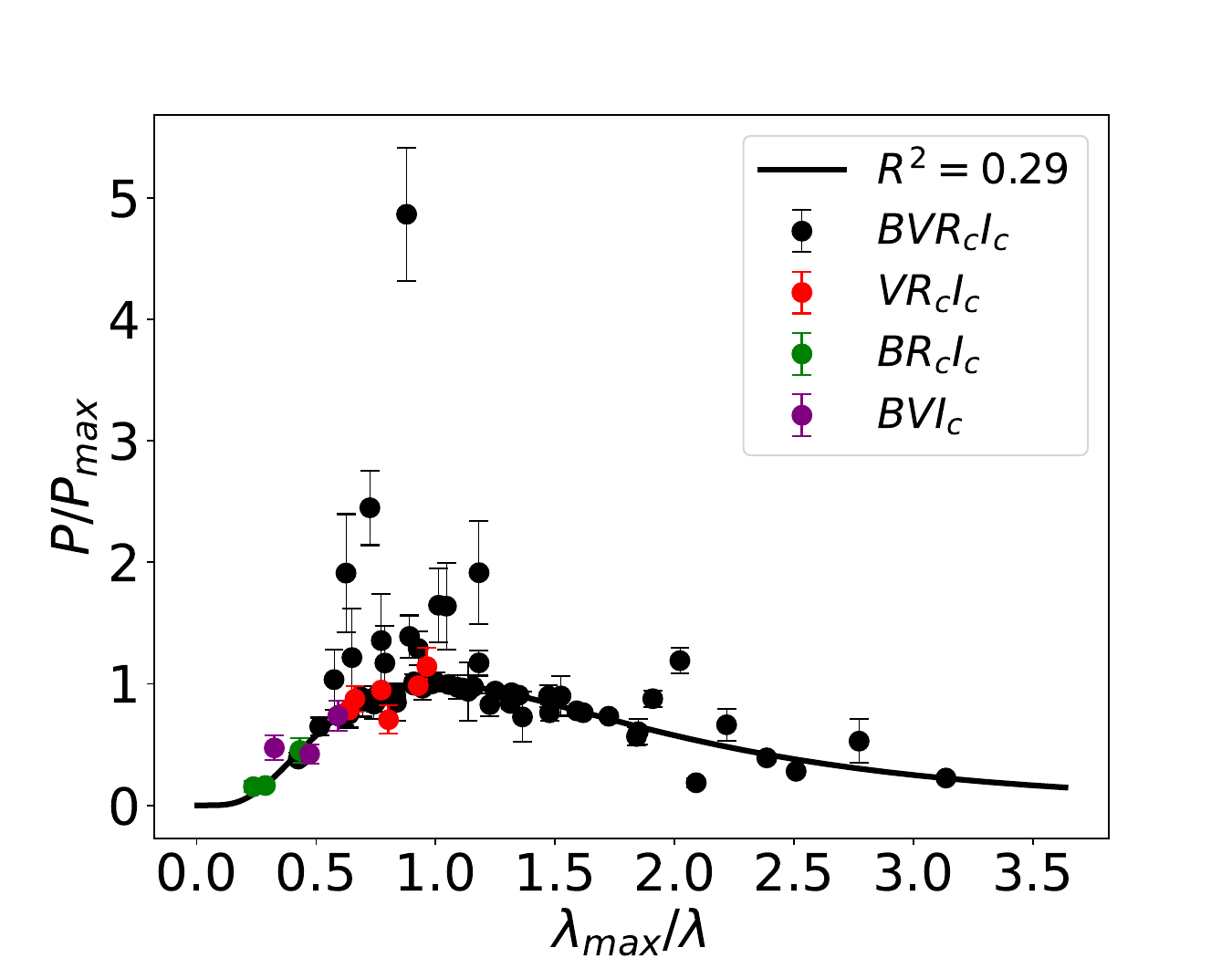}
\caption{The plot of $P/P_{\text{max}}$ against $\lambda_{\text{max}}/\lambda$ showing the normalized polarization-wavelength dependence for 19 stars towards AFGL 6366S in the $BV(RI)_c$ wavelength bands. The black points denote stars satisfying the $P/e_p > 3$ criterion in all four Optical bands. The colored points indicate stars meeting this criterion in three bands only: red ($VR_cI_c$), green ($BR_cI_c$), and purple ($BVI_c$). These are also indicated in the legend.  The uncertainties for the measurements of $P/P_{\text{max}}$ are represented by the error bars and the solid curve reflects the Serkowski polarization relation for general diffuse ISM. The co-efficient of determination of fit ($R^2$) between the observed data and serkowski relation is shown in the top right corner.}
\label{Fig:ser_opt}    	
\end{figure}

We modeled the observed $BVR_cI_c$ polarization (listed in Table \ref{Tab:Opt_pol}) using the standard Serkowski law (Eq. \ref{eq:ser_law}) adopting $k = 1.15$. To ensure the accuracy of the fit, only stars with reliable polarization measurements - defined as those with $P/e_p > 3$ in at least three wavelength bands - were considered. Based on this criterion, 19 out of the 31 stars were included in the fit. Fig. \ref{Fig:ser_opt} illustrates the normalized polarization wavelength dependence for these 19 stars. The black coloured points correspond to stars that satisfy the reliability criterion across all four Optical bands. The points in other colours represent stars satisfying the criterion in only three bands: red for $VR_cI_c$, green for $BR_cI_c$, and purple for $BVI_c$. The black solid curve represents the Serkowski model for the general ISM. The coefficient of determination of fit ($R^2$) between the observed data and the theoretical model is $\approx 0.29$. This indicates a relatively poor fit, which might be due to a substantial contribution from intrinsic polarization. 

Table \ref{tab:ser_law} presents the best-fit values of the Serkowski parameters, $P_{\text{max}}$ and $\lambda_{\text{max}}$, for the 19 stars that were included in the fit. The Table also includes the unit weight error of the fit ($\sigma_1$) for each star. This metric quantifies the deviation of the observed polarization from the Serkowski relation \citep{orsatti_2006}. 




When the observed polarization arises predominantly from interstellar dust, the wavelength dependence should comply with the Serkowski law. Empirical studies show that, in such cases, the corresponding $\sigma_1$ value typically remains below 1.6 \citep{medhi_2007, Medhi_2008, Medhi_2010, eswaraiah_2011, Biswas_2024}. A higher $\sigma_1$ value suggests an additional intrinsic polarization component \citep{Biswas_2024}. Using this criterion, 15 stars (\#5, \#18, \#24, \#30, \#38, \#43 \#45, \#49, \#51, \#54, \#55, \#56, \#58, \#62 and \#64) out of the 19, which were included in the fit, are identified as potential candidates of intrinsic polarization. The parameter $\lambda_{\text{max}}$ also serves as an indicator of intrinsic polarization. Stars with $\lambda_{\text{max}}$ values significantly lower than the typical interstellar average \citep[$0.545 \pm 0.04 \ \mu m$;][]{serkowski_1975} are likely to possess an intrinsic polarization component \citep{eswaraiah_2011}. In this context, Stars \#55, \#56, and \#64 present strong signatures of intrinsic polarization, indicated by their significantly low $\lambda_{\text{max}}$ values together with $\sigma_1 > 1.6$. 

The large number of stars showing intrinsic polarization is expected. AFGL 6366S is a young star-forming region with many pre-main-sequence objects, including Class I/II protostars and T Tauri stars \citep{2025MNRAS.541.1557B}. Such young stellar objects (YSOs) often possess circumstellar disks or envelopes that can produce significant intrinsic polarization, contributing to the observed deviations from the Serkowski law \citep{2008AJ....136..621K,2009A&A...501..595P}. 

\subsubsection{JHK$_s$ Polarization}
\label{sec: wave_dep_nir}

To model the $JHK_s$ polarization measurements (listed in Table \ref{Tab:NIR_pol}), we first applied the modified Serkowski law by adopting $k = (-0.10 \pm 0.05) + (1.86 \pm 0.09)\lambda_{\text{max}}$. Among the total 64 stars in the sample, 53 satisfied the condition $P/e_p > 3$ in all the three NIR bands. Individual fits were performed for each of these stars. However, many of these fits produced implausibly large uncertainties in $P_{\text{max}}$ and $\lambda_{\text{max}}$. Only 10 of the 53 stars exhibited parameter uncertainties smaller than one-third of the best  fit values. These outcomes imply that, within the constraints of the current NIR dataset, the modified Serkowski law may not entirely represent the polarization behaviour.

\begin{figure*}
\begin{subfigure}{\columnwidth}\hspace{0.5cm}
\includegraphics[width=0.8\columnwidth]{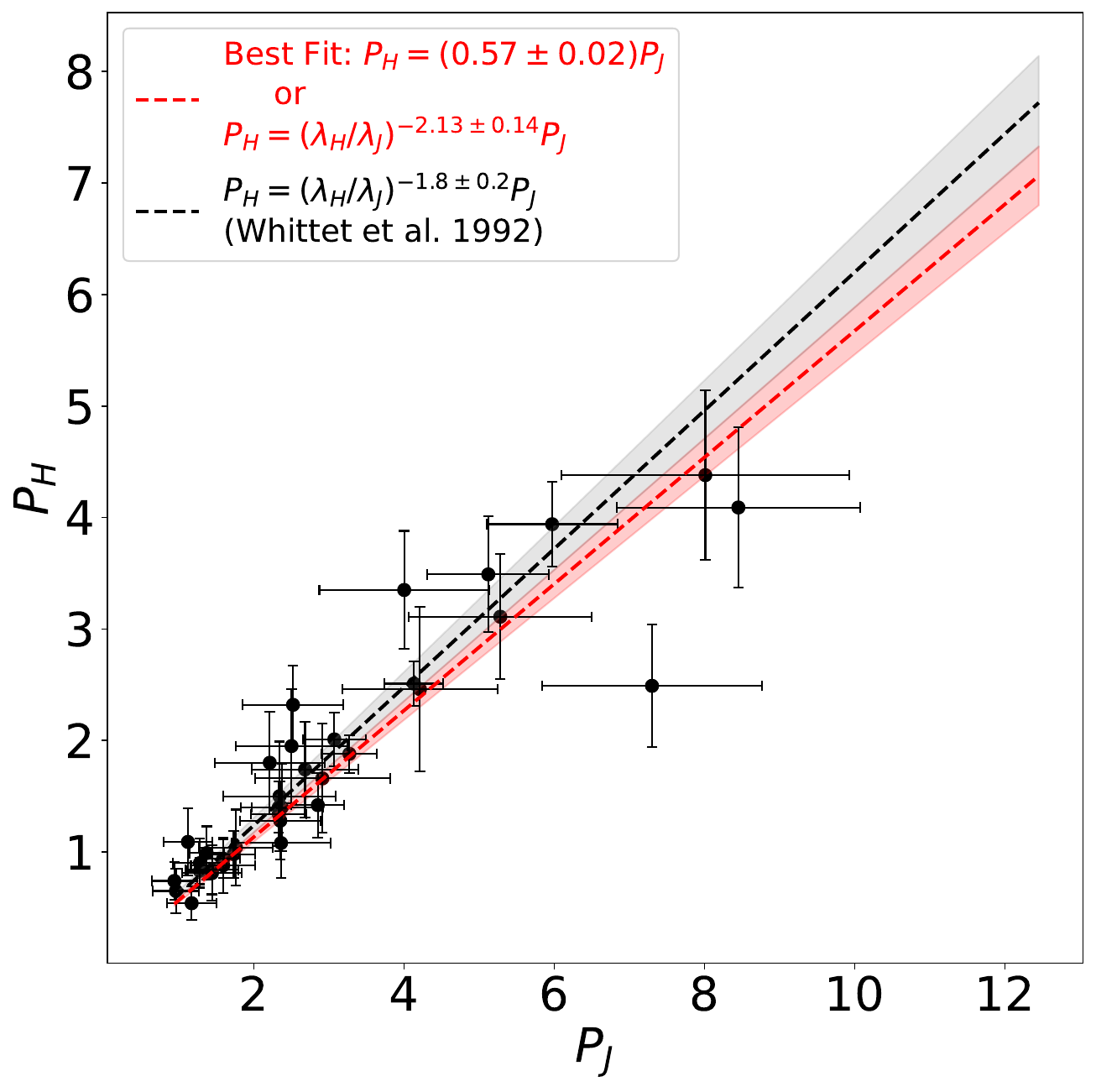}
\caption{}
\label{Fig:NIR_pol_JH_depend}
\end{subfigure}
\begin{subfigure}{\columnwidth}\hspace{0.5cm}
\includegraphics[width=0.8\columnwidth]{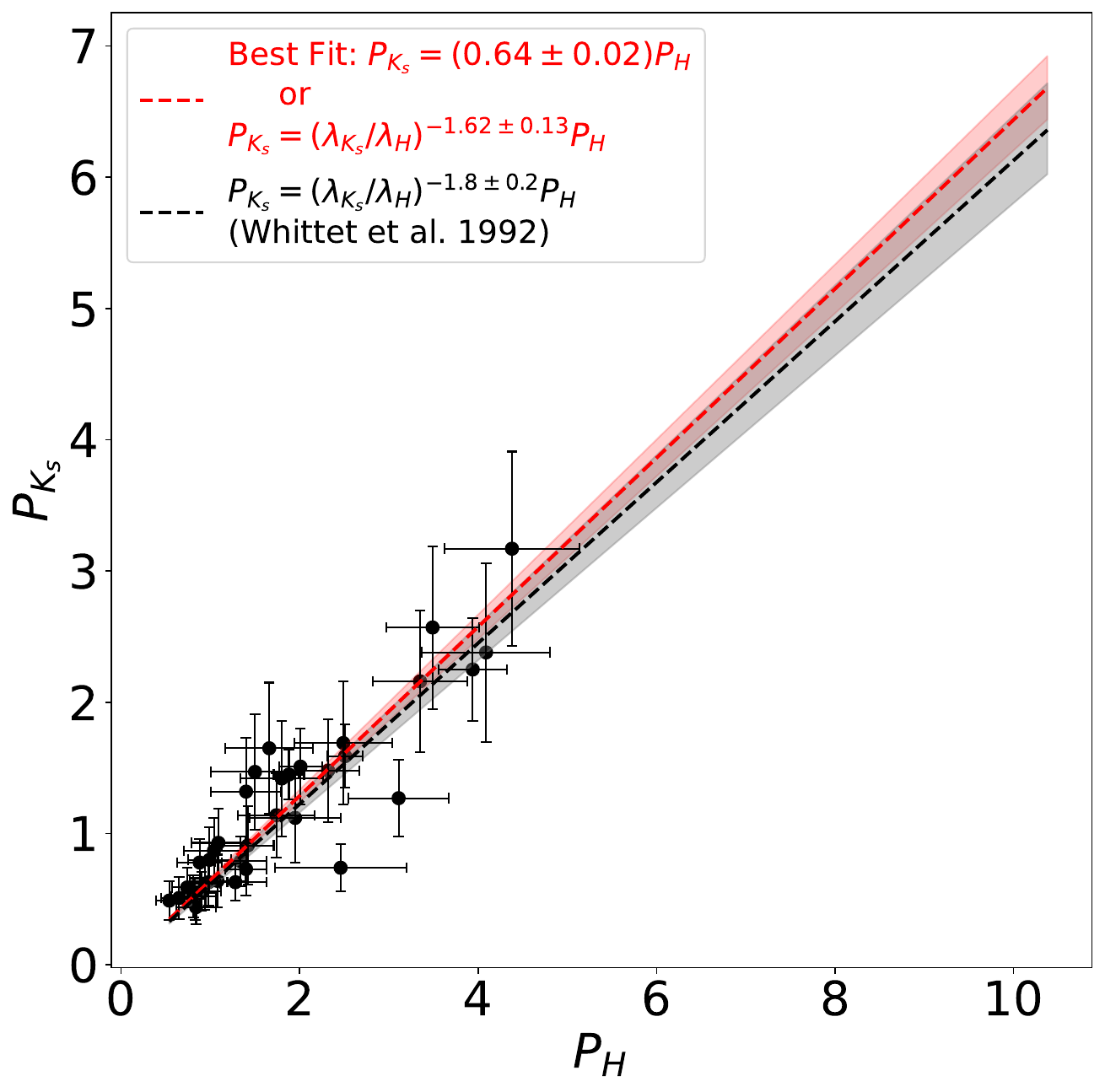}
\caption{}
\label{Fig:NIR_pol_KH_depend}
\end{subfigure}
\caption{Plots of (a) $P_H$ versus $P_J$ and (b) $P_H$ versus $P_{K_s}$ for 36 sources (black filled circles) satisfying $P/e_P > 3$ in all three $JHK_s$ bands and exhibiting a monotonic decrease in polarization from 1.25 to 2.14 $\mu$m. Error bars represent the corresponding polarization uncertainties. The red dashed lines and their corresponding shaded regions denote the best-fit relations and their 1$\sigma$ uncertainties, while the black dashed line and gray shaded region represent the \citep{whittet_1992} model. These relations are indicated by the corresponding colors in the legends of both panels.}
\label{Fig:NIR_pol_depend}
\end{figure*}

In view of this, we examined the power-law wavelength dependence of the $JHK_s$ polarization. We modeled the $P_J$–$P_H$ and $P_H$–$P_{K_s}$ correlations, shown in Figs. \ref{Fig:NIR_pol_JH_depend} and \ref{Fig:NIR_pol_KH_depend}, using a simple linear fit. To ensure a robust fit, we again considered the sources satisfying \textit{P}/$e_P >$ 3 in all three NIR bands. We further restricted the sample to those exhibiting the expected monotonic decrease in polarization with increasing wavelength from 1.25 to 2.14 $\mu m$ \citep[e.g.,][]{Nakajima_2007, Hatano_2013}. These criteria provided a subset of 36 stars from the full sample of 64. 

The red dashed lines in Figs. \ref{Fig:NIR_pol_JH_depend} and \ref{Fig:NIR_pol_KH_depend} show the best-fit relations, with slopes $\left<P_H/P_J\right>$ = 0.57 $\pm$ 0.02 and $\left<P_{K_s}/P_H\right>$ = 0.64 $\pm$ 0.02; the corresponding 1$\sigma$ confidence intervals are indicated by the red shaded regions. In the power law form these relations can be expressed as $P_H = (\lambda_H / \lambda_J)^{-2.13~\pm~0.14} P_J$ and $P_{K_s} = (\lambda_{K_s} / \lambda_H)^{-1.62~\pm~0.13} P_H$, respectively. The derived values are consistent, within uncertainties, with the \cite{whittet_1992} model, shown in the figures by the black dashed lines and gray 1$\sigma$ confidence regions.





\begin{figure}\hspace{0.5cm}
\includegraphics[width = 0.8\columnwidth]{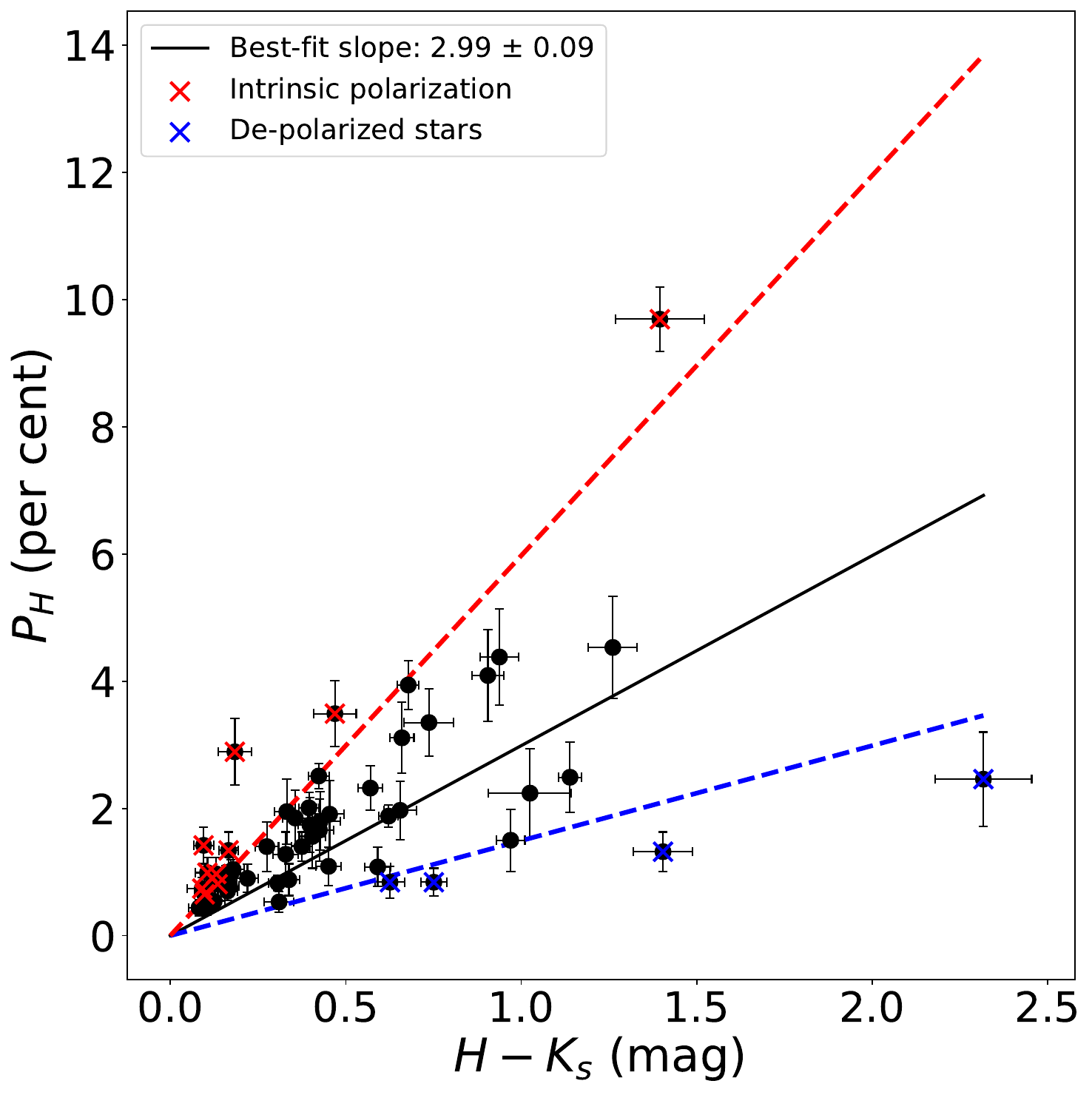}
\caption{$P_H$ versus $(H-K_s)$ diagram for 53 stars (black filled circles) from Table \ref{Tab:NIR_pol}, satisfying the $P/e_p > 3$ criterion and having high-quality 2MASS photometry. The horizontal and vertical error bars represent uncertainties in color and polarization, respectively. The black solid line represents the best-fit linear relation. The best-fit slope is shown in legend at the top left corner. The dashed red and blue lines correspond to slopes twice and half of the best-fit slope, respectively. Probable sources of intrinsic polarization and stars exhibiting depolarization are marked using red and blue crosses, respectively.}
\label{Fig:nir_in_pol}    	
\end{figure}

Polarization–color diagrams (PCDs) are widely used in NIR polarimetric studies for separating interstellar and intrinsic stellar polarization \citep[e.g.,][]{2008AJ....136..621K,2010ApJ...716..299S}. In this context, we constructed the $P_H$ versus $(H-K_s)$ PCD shown in Fig. \ref{Fig:nir_in_pol}, adopting $(H-K_s)$ colors from the 2MASS PSC \citep{Cutri_2003, Skrutskie_2006}. Out of the 64 NIR polarimetric sources, 53 satisfying the $P/e_p > 3$ criterion and having high-quality 2MASS photometry were included in the analysis (as described in Sec. \ref{Sec:2mass_data}). For simplicity, we focus only on the $H$-band polarization. However, almost similar trends are observed with $J$ and $K_s$ band polarization data.

To characterize the overall interstellar polarization trend, we fitted a linear relation to the PCD, shown as the solid black line in Fig.~\ref{Fig:nir_in_pol}. The best-fit slope is $P_H/(H-K_s) = 2.99 \pm 0.09$ per cent mag$^{-1}$. Following the methodology adopted in earlier polarization studies \citep[e.g.,][]{2015ApJ...798...60K, 2017ApJ...850..195E}, we then draw two additional reference lines with slopes equal to twice (slope $\sim$ 5.98) and half (slope $\sim$ 1.50) of the best-fit slope. These are shown in Fig.~\ref{Fig:nir_in_pol} as dashed red and blue lines. They provide a practical framework for identifying stars that deviate significantly from the average interstellar trend of our target region. Stars above the upper red line are flagged as probable intrinsic polarization sources. Stars below the lower blue line are classified as under-polarized. These are likely affected by depolarization. Similar criteria have been used extensively in earlier NIR polarization works \citep[e.g.,][]{2008AJ....136..621K,2010ApJ...716..299S, 2015ApJ...798...60K, 2017ApJ...850..195E}.

As an independent check, we also examined an alternative criterion proposed by \citet{2008AJ....136..621K}. They adopted the polarization efficiency of BN (one of the most highly polarized dichroic sources known with $P_H \sim 31$ per cent) as an empirical upper limit for interstellar polarization. It corresponds to $P_H/(H-K_s) \sim 8.2$ per cent mag$^{-1}$. Stars above this limit were considered intrinsically polarized. Applying this criterion to our data, provides results which are fairly consistent with our former method. However, it identifies slightly fewer intrinsic polarization candidates. In comparison, our former approach adopts a lower interstellar polarization limit of $P_H/(H-K_s) \sim 5.98$ per cent mag$^{-1}$, thereby imposing a more stringent threshold that is likely to be more robust. Moreover, because the maximum polarization efficiency can vary from region to region, the BN-based limit may not be directly applicable to the AFGL 6366S region. This further supports our use of a region-specific threshold. Accordingly, we rely only on the former method to identify sources exhibiting intrinsic polarization in our NIR dataset.

Applying this approach, we identified 10 stars (with IDs \#8, \#16, \#17, \#27, \#45, \#48, \#50, \#51, \#54, \#63; marked with red cross in Fig. \ref{Fig:nir_in_pol}) as probable sources of intrinsic polarization. Three of these  (\#45, \#51, \#54) were also identified as intrinsically polarized based on the Optical polarization data. Therefore, considering the Optical and NIR results together, it is inferred that a total of 22 stars in the region may possess polarization of non-interstellar origin.

Additionally, four stars (\#9, \#15, \#22, \#61; marked with blue cross in Fig. \ref{Fig:nir_in_pol}) were found to exhibit signs of depolarization. The presence of these depolarized stars indicate towards a complex and less coherent magnetic field geometry or reduced grain alignment efficiency in some of the regions towards AFGL 6366S \citep{2017ApJ...850..195E}.

\subsection{Polarization efficiency}

The polarization efficiency of interstellar dust grains is defined as the degree of polarization produced per unit extinction or reddening. It primarily depends on factors such as the magnetic field orientation and strength, degree of alignment of the dust grains and the amount of depolarization \citep{Medhi_2008, Voshchinnikov_2012, 2022MNRAS.513.4899S}. Thus, various galactic environments, e.g. diffuse ISM, molecular clouds,  and star-forming regions, etc., show different behaviour of polarizing efficiency. In the diffuse ISM, the empirical upper limit of the maximum attainable polarization is expressed as \citep{1956ApJS....2..389H, serkowski_1973, serkowski_1975}:

\begin{equation}
	{P_{\rm max} < 3A_{V} \simeq 3R_{V} \times E(B-V)}
	\label{Eq:pol_eff}
\end{equation}

where, $A_{V}$ denotes the visual extinction, $E(B-V)$ denotes the color excess and $R_{V} = A_{V}/E(B-V)$ is the total-to-selective extinction ratio.

\begin{figure*}
\begin{subfigure}{\columnwidth}
\includegraphics[width=\columnwidth]{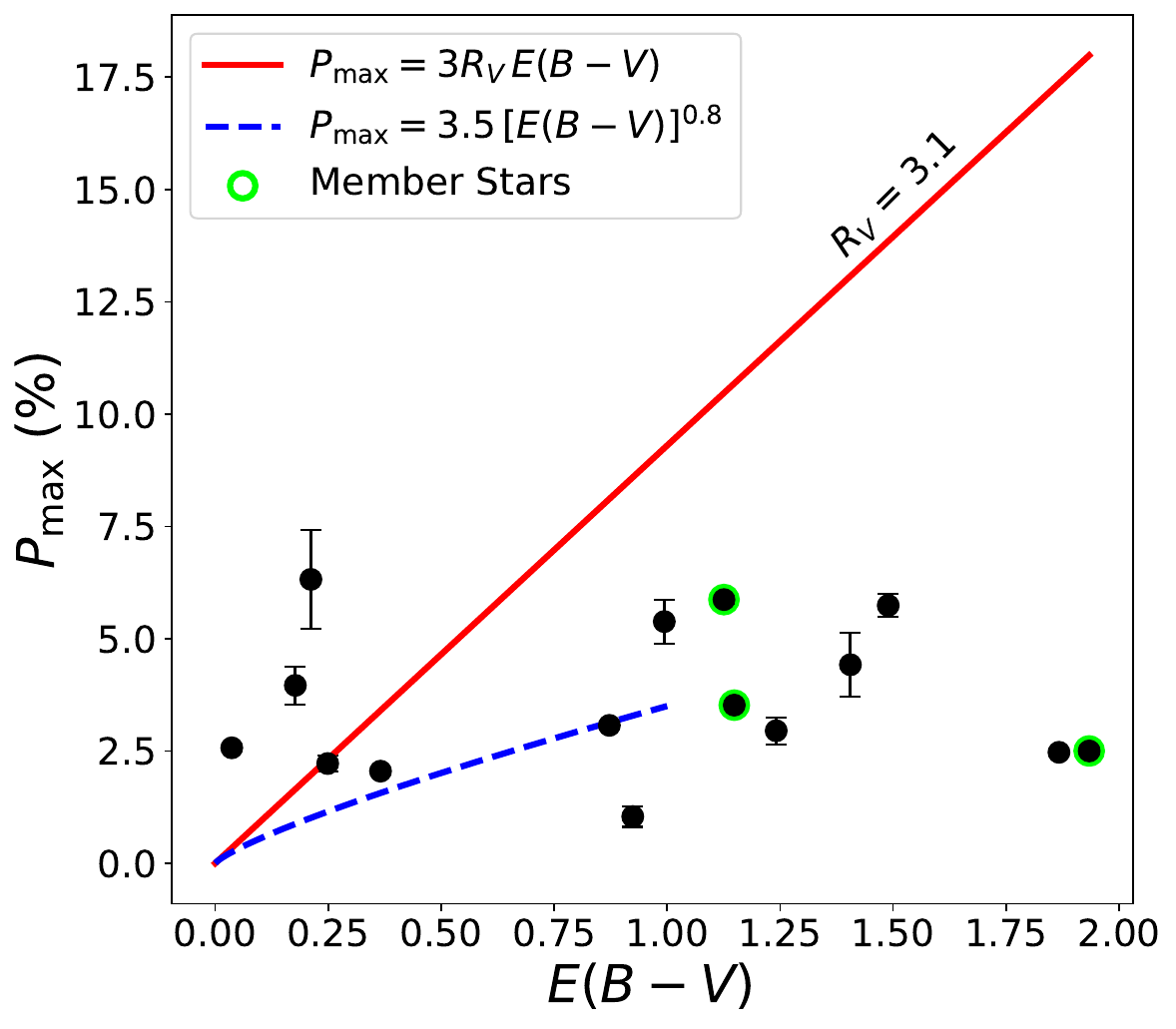}
\caption{}
\label{Fig:pol_eff1}
\end{subfigure}
\begin{subfigure}{\columnwidth}
\includegraphics[width=\columnwidth]{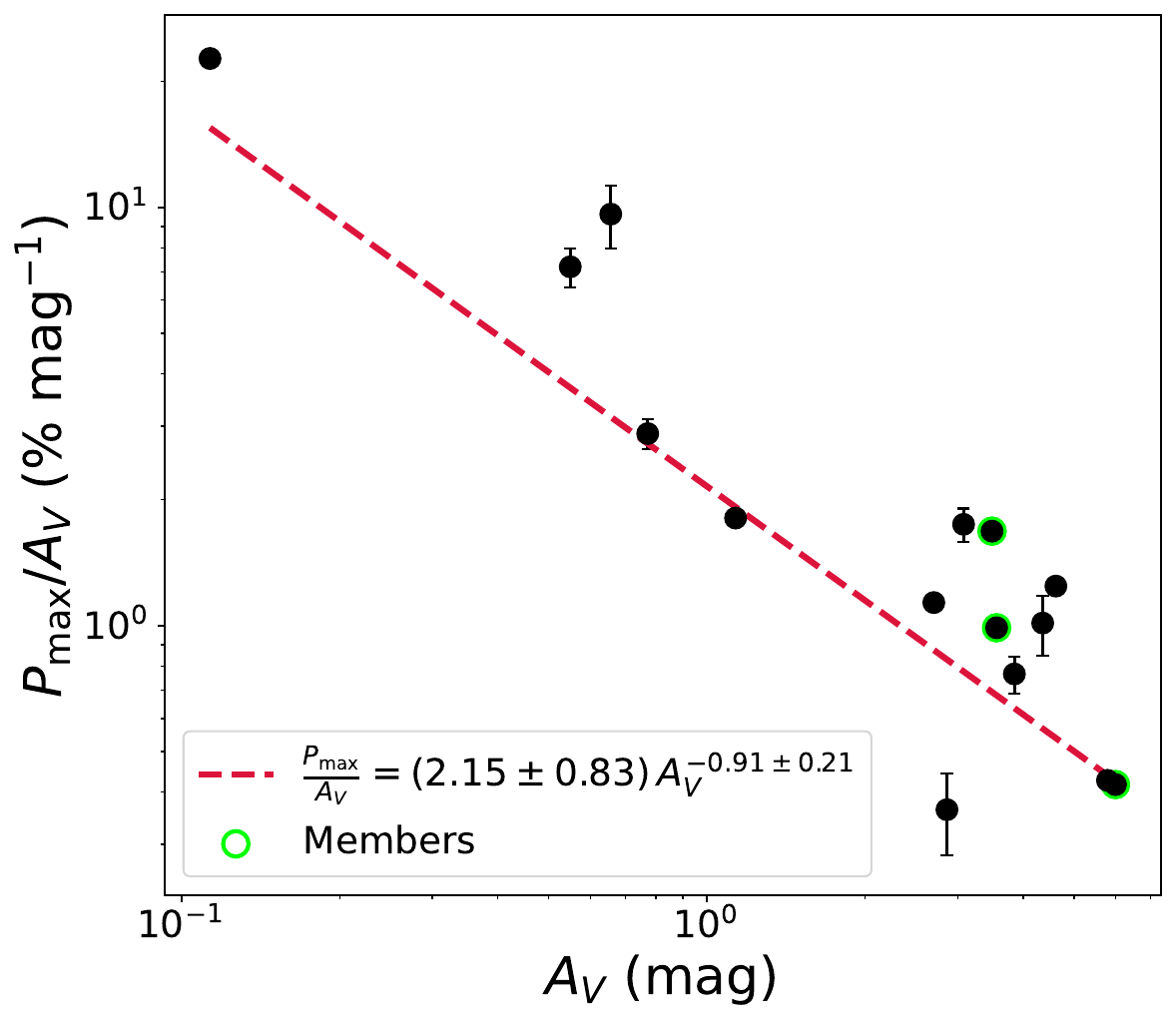}
\caption{}
\label{Fig:pol_eff2}
\end{subfigure}
\caption{(a) Variation of $P_{\text{max}}$ with $E(B-V)$ for 15 stars (filled black circles) out of the 19 stars listed in Table \ref{tab:ser_law} with reliable color excess estimates. The red line denotes the empirical upper limit of polarization efficiency for $R_V = 3.1$, while the blue curve represents the average Galactic relation. (b) $P_{\text{max}}/A_V$ as a function of $A_V$ for the same 15 stars. Error bars in both panels represent uncertainties in $P_{\text{max}}$ (uncertainties in $A_V$ were not available). The red dashed line shows the best-fit power-law relation in logarithmic scale. The best fit relation is also indicated in the legend. In both panels, green circles indicate member stars of AFGL 6366S.}
\label{Fig:pol_eff}
\end{figure*}

To estimate the polarizing efficiencies of the dust grains towards AFGL 6366S, we constructed the polarization efficiency diagram (shown in Fig. \ref{Fig:pol_eff1}) using 15 out of the 19 stars listed in Table \ref{tab:ser_law} for which reliable color excesses could be derived. The $E(B-V)$ values for these sources were obtained from their $A_V$ estimates reported in \cite{2022A&A...658A..91A}, adopting $R_V = 3.1$ for the conversion. Uncertainties in $A_V$ were not available. The remaining 4 stars lacked $A_V$ estimates and thus could not be included in the diagram. 

The red straight line in Fig. \ref{Fig:pol_eff1} corresponds to the emperical upper limit (Eq. \ref{Eq:pol_eff}) for $R_V$ = 3.1. The blue dashed curve shows the average polarizing efficiency in the Galaxy, expressed as \citep{2002ApJ...564..762F}:

\begin{equation}
	P_{\rm max} = 3.5 \times E(B-V)^{0.8}, \quad \text{for } E(B-V) < 1.0~\text{mag}.
\end{equation}

It can be clearly seen in Fig. \ref{Fig:pol_eff1} that the majority of stars in the selected sample exhibit polarization efficiencies much below the empirical maximum. These also include 3 cluster members which are shown using the green circles. Several stars fall even below the average Galactic efficiency curve. This suggests a relatively low alignment efficiency of dust grains in this region. Only 3 stars (\#43, \#45, and \#55) show polarization efficiencies exceeding the maximum expected value. However, all three of them have $\sigma_1 > 1.6$ (see Table~\ref{tab:ser_law}). So, this could be due to the intrinsic nature of polarization in these stars.

Dust grain alignment within a region can be evaluated by the variation of polarization efficiency with extinction. Several studies \citep[e.g.,][]{2018MNRAS.475.5535I, 2021AJ....161..149S} have reported a decrease in polarization efficiency with increasing extinction. It often follows a power-law relation of the form: $P_{\text{max}}/A_V = \beta(A_V)^\alpha$. In the diffuse ISM, the power-law index $\alpha$ is typically close to $-0.5$ \citep{2022MNRAS.513.4899S}. In dense molecular or dust-embedded regions, $\alpha$ may approach $-1$ \citep{2015AJ....149...31J, 2014A&A...569L...1A}.


For the AFGL~6366S region, the power-law fit to the observed polarization efficiency ($P_{\rm max}/A_V$) as a function of $A_V$ is shown in Fig.~\ref{Fig:pol_eff2}. The logarithmic scale has been used for better visualization. The best-fit parameters were estimated as $\beta = 2.15 \pm 0.83$ and $\alpha = -0.91 \pm 0.21$, using the bootstrap resampling technique \citep{efron1992bootstrap, efron1994introduction}.



The derived value of $\alpha = -0.91 \pm 0.21$, indicates a steep decline in polarization efficiency with increasing extinction. This again suggests significant depolarization effects at higher extinction regions.

\subsection{Column density and dust temperature}
\label{Sec: cdens_dtemp}
\begin{figure*}
\begin{subfigure}{\columnwidth}
\includegraphics[width=\columnwidth]{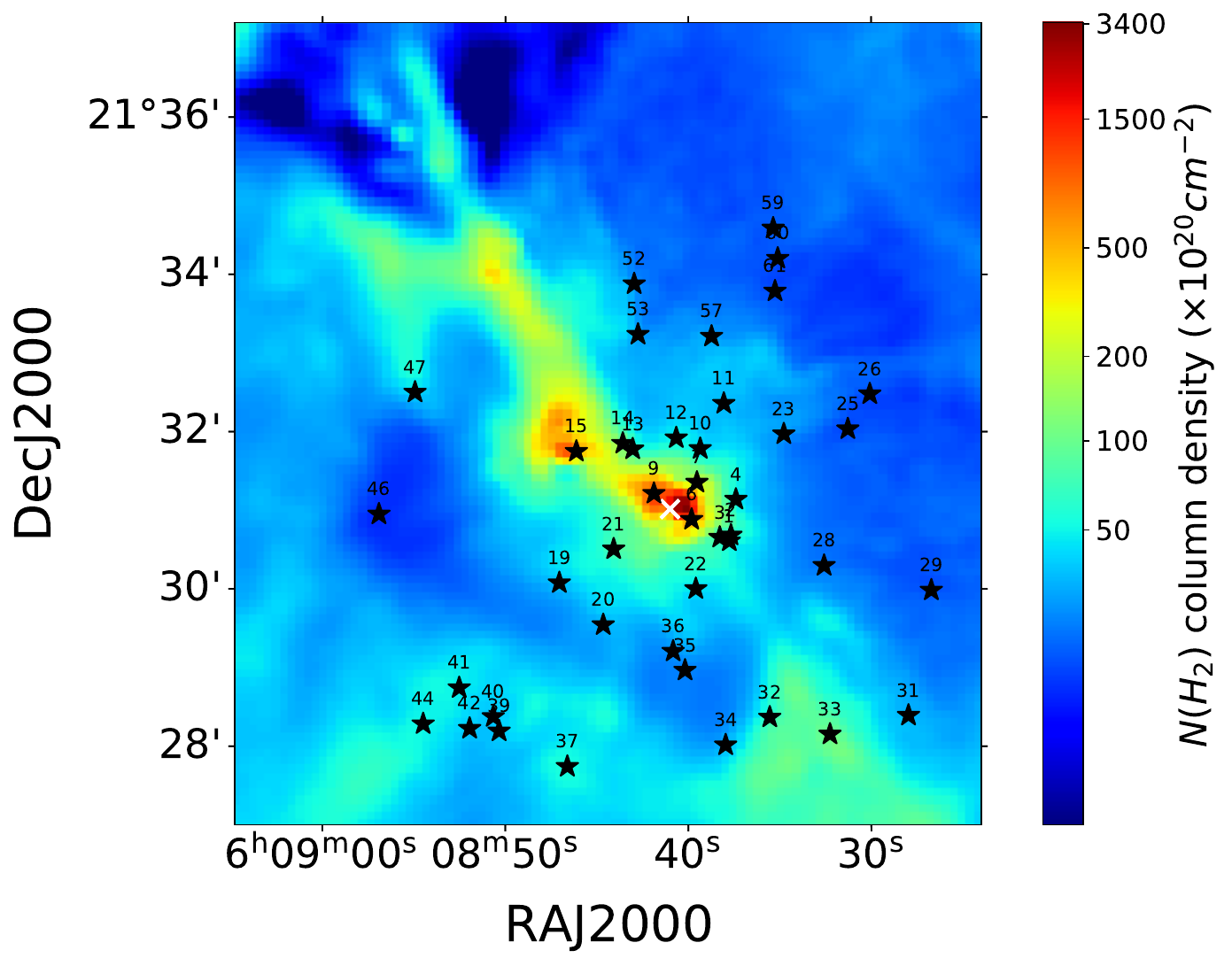}
\caption{}
\label{Fig:cdens_map}
\end{subfigure}
\begin{subfigure}{\columnwidth}
\includegraphics[width=\columnwidth]{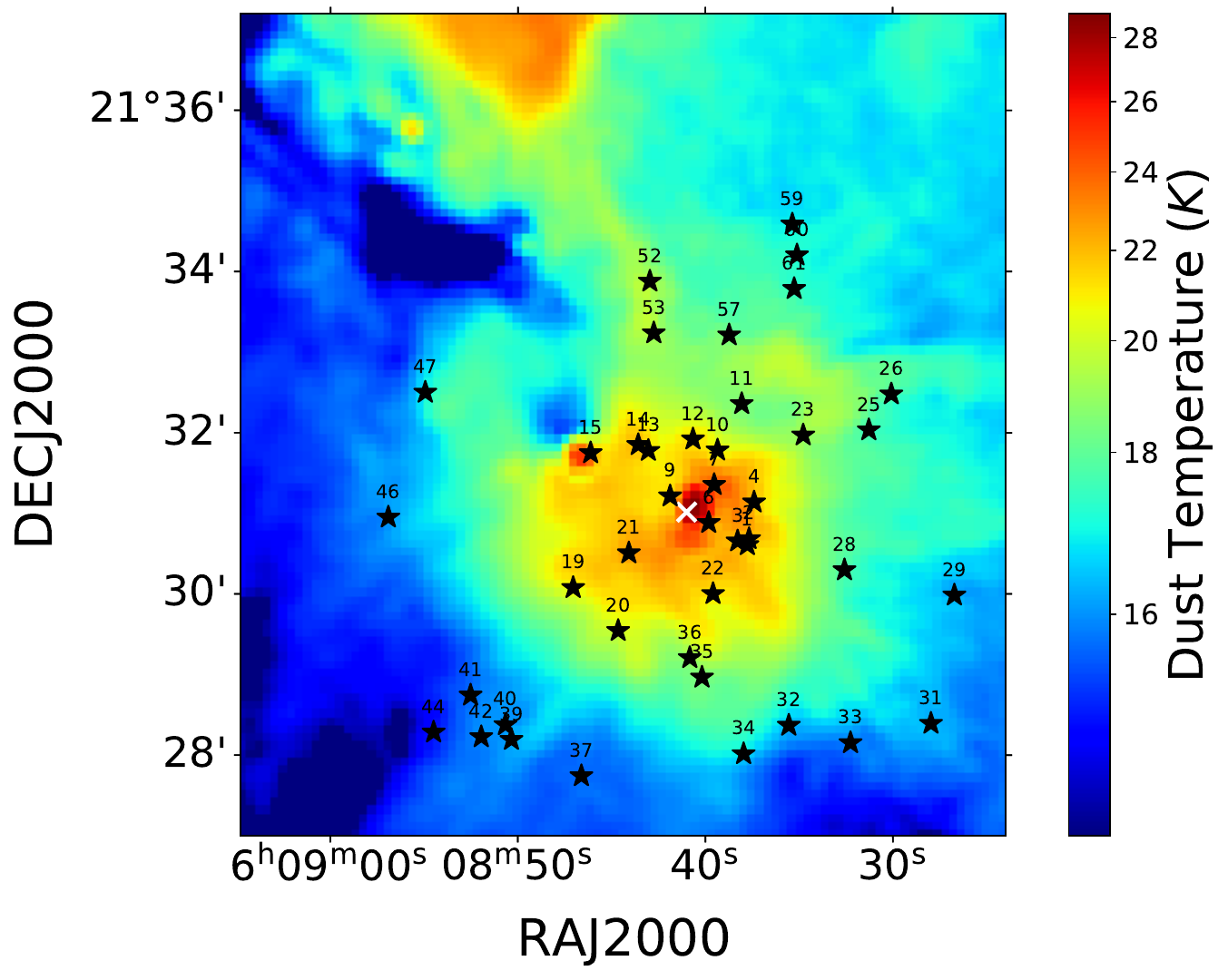}
\caption{}
\label{Fig:dtemp_map}
\end{subfigure}
\caption{(a) Column density map and (b) dust temperature map of the AFGL 6366S region, each covering an area of $10 \times 10$ arcmin$^2$. In both panels, star symbols indicate the polarimetrically observed stars, excluding those exhibiting possible intrinsic polarization. The corresponding star IDs, as listed in Table \ref{Tab:NIR_pol}, are annotated. The white cross mark denotes the centre of the cluster AFGL 6366S.}
\label{Fig:cdens_dtemp_map}
\end{figure*}

We investigated the spatial distributions of the molecular hydrogen column density ($\mathrm N(\mathrm H_2)$) and dust temperature ($T_d$) across the AFGL 6366S region, using 2D column density and temperature maps obtained from the \href{http://www.astro.cardiff.ac.uk/research/ViaLactea/}{PPMAP results archive}. These were generated by applying the Point Process MAPping (PPMAP) technique \citep{2015MNRAS.454.4282M, 2017MNRAS.471.2730M, 2020ascl.soft04008M} to the Hi-GAL survey data \citep{2010PASP..122..314M}. These are presented in Figs.~\ref{Fig:cdens_map} and \ref{Fig:dtemp_map}. The black star symbols represent our polarimetrically observed stars. The stars exhibiting possible intrinsic polarization were excluded.

Figs.~\ref{Fig:cdens_map} and \ref{Fig:dtemp_map} show that both column density and dust temperature attain their maximum values in the central region of the AFGL 6366S cluster. In both the figures, the cluster centre is marked by a white cross. The peak $\mathrm N(\mathrm H_2)$ and $\mathrm T_d$ values are estimated as $\sim$ $3.4 \times 10^{23}$~cm$^{-2}$ and 28.8~K, respectively. These are measured using the  \href{https://sites.google.com/cfa.harvard.edu/saoimageds9/download}{SAOImage DS9} software. The elevated dust temperature towards the cluster center, relative to surrounding regions, likely arises from local heating caused by the embedded young stellar objects (YSOs) within the cluster core. Such localized heating, together with the presence of dense molecular material, is a well-known signature of active or recent star formation events \citep{2020MNRAS.496.5201M}. 

A prominent filamentary structure is evident in Fig.~\ref{Fig:cdens_map}. This feature was also identified in \href{https://academic.oup.com/mnras/article/541/2/1557/8173845}{Paper~I}
 and in earlier studies \citep[e.g.,][]{shimoikura_2013, maity2023MNRAS.523.5388M}. The long axis of the filament is oriented at an angle of $\theta_{\rm fil} \approx 216^\circ$ measured from north towards east. To estimate the mean column density of the filamentary structure, it was interactively outlined using the \texttt{PolygonSelector} tool in \texttt{\href{https://matplotlib.org/}{Matplotlib}} python pacakage. A mask was then generated to extract the enclosed pixel values. The resulting mean column density of the filament was estimated as $(4.1 \pm 0.3) \times10^{22}$ cm$^{-2}$. The error was computed as $\sigma / \sqrt{N}$, where $\sigma$ is the standard deviation of the sampled pixels and $N$ is the number of pixels selected.

\subsection{Grain alignment efficiency and magnetic tangling}
\label{sec:ga}

The efficiency of grain alignment is strongly affected by local physical conditions, including the strength of the radiation field (often reflected by the dust temperature) and the surrounding gas density \citep{2025ApJ...981..128P}. To investigate the dependence of interstellar polarization on column density and temperature, we sampled the $N(\mathrm{H}_2)$ and $T_d$ values at the locations of the observed stars, from the corresponding maps shown in Fig.~\ref{Fig:cdens_dtemp_map}. This was done using the \texttt{\href{https://matplotlib.org/}{Matplotlib}} python pacakage. For each star, a square aperture of $2 \times 2$ pixel$^2$ was defined, centered on the star's pixel coordinates. Each pixel represents $\sim$ 6 arcsec on the sky, which is the effective angular resolution of the column density and temperature maps. The chosen aperture thus samples an area roughly 5 times larger than the typical stellar FWHM ($\sim$1.7 arcsec; see Table \ref{tab:obs_summary_nir}) in our observed NIR data. This ensures that the measurements reflect the local interstellar environment around each star. All pixels within this aperture were used to compute the mean values of column density and temperature as well as the associated uncertainties. These values were then assigned to the corresponding stars and adopted as representative measurements of their local column density and dust temperature.

\begin{figure}\hspace{0.5cm}
\includegraphics[width=0.9\columnwidth]{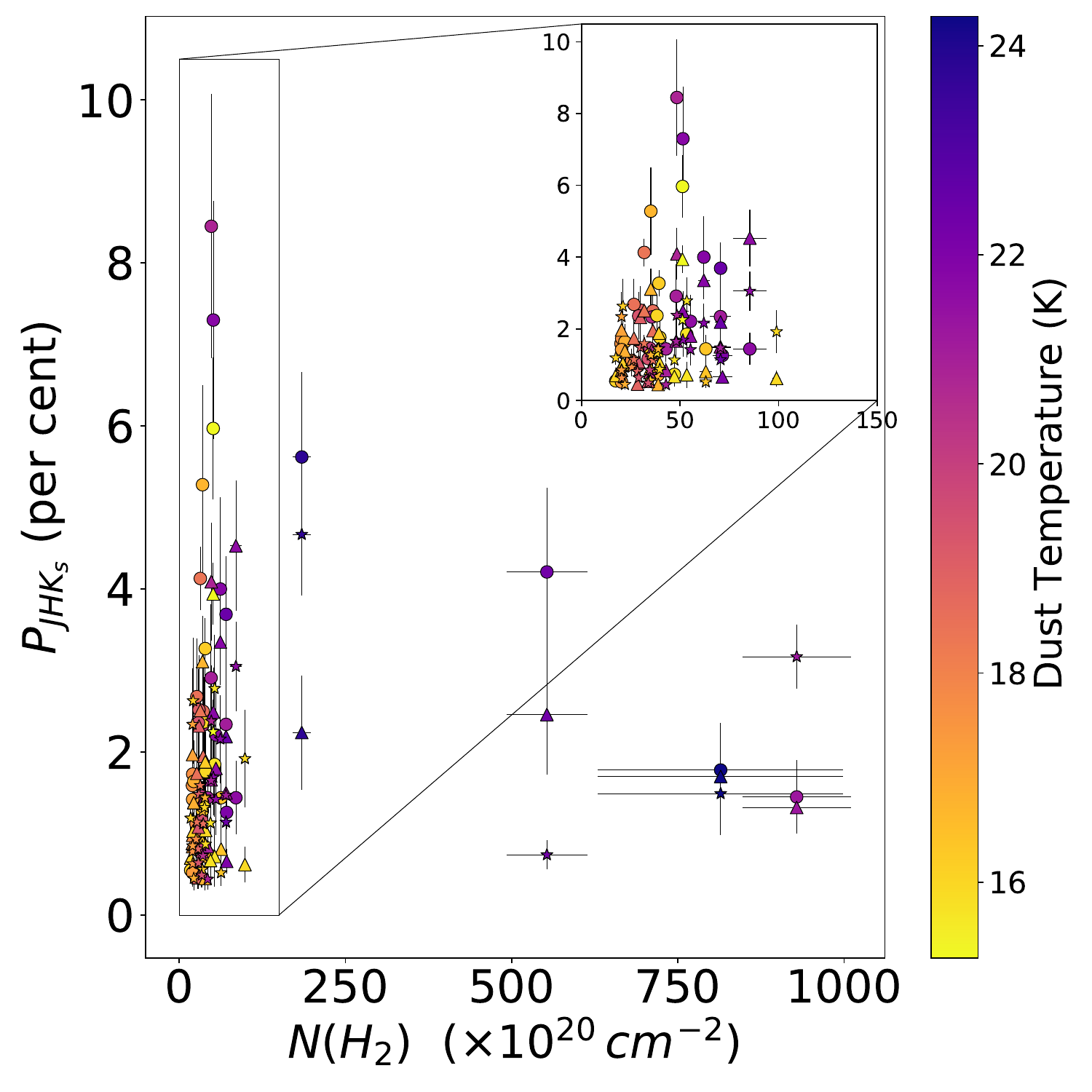}
\caption{The variation of $P_{JH K_s}$ with $N(H_2)$ for the stars shown in Fig.~\ref{Fig:cdens_dtemp_map}. Circles, triangles, and star symbols represent $P_J$, $P_H$, and $P_{K_s}$, respectively. Each data point is color-coded based on the corresponding dust temperature ($T_d$). To improve visualization, an inset in the top-right corner highlights the data within the $N(H_2)$ range of [0, 180]}
\label{Fig:cdens_temp_pol}
\end{figure}

Fig.~\ref{Fig:cdens_temp_pol} shows how the $J$, $H$, and $K_s$ polarization ($P_{JHK_s}$) varies with $\mathrm{N}(\mathrm{H}_2)$. Each data point (representing an observed star) is color-coded with its associated dust temperature. It can be seen clearly that the polarization percentage initially increases with the increasing column density and/or temperature. It reaches a maximum value of 8.45 per cent in the $J$ band, 4.53 per cent in the $H$ band and 4.67 per cent in the $K_s$ band. These peak values correspond to $N(\mathrm{H}_2) \approx 4.8 \times 10^{21}$ cm$^{-2}$ and $T_d \approx 21$ K for $J$ band, $ 8.5 \times 10^{21}$ cm$^{-2}$ and $ 22$ K for $H$ band, and $1.8 \times 10^{22}$ cm$^{-2}$ and $ 24$ K for $K_s$ band. Beyond these points, the polarization in all three bands are found to decline. Similar decreases in polarization for $T_{\rm d} > 19-25 K$ have also been reported in previous observational studies of molecular clouds regions \citep[e.g.,][]{2019ApJ...882..113S,2021ApJ...906..115T,2025ApJ...981..128P}. This phenomenon is commonly referred to as the polarization hole \citep{2019FrASS...6...15P}. This decrease in polarization is generally attributed to a reduced grain-alignment efficiency and/or magnetic field tangling along the line of sight \citep{2025ApJ...981..128P}. Magnetic field tangling arises when field lines within the observed region deviate from a uniform orientation. This might be due to turbulence, gravitational collapse or stellar feedback. Consequently, polarized light from regions with differing field orientations partially cancels along the line of sight. This reduces the net degree of polarization \citep{2018ApJ...857L..10C}.

Nevertheless, the initial increase in polarization with the increasing dust temperature is consistent with the Radiative Torque Alignment (RAT-A) theory \citep{1976Ap&SS..43..291D}. According to this theory, non-spherical dust grains exposed to an anisotropic radiation field, experience radiative torques that spin them up to suprathermal velocities. This aligns their short axes parallel to the local magnetic field \citep{2007MNRAS.378..910L}. Since higher dust temperatures typically correspond to stronger radiation fields, grains in these environments experience more efficient alignment. Hence, they cause higher polarization. 

The subsequent decline of polarization at still higher dust temperatures cannot be accounted for by RAT-A. However, this decline is well explained by Radiative Torque disruption (RAT-D) mechanism \citep{2019NatAs...3..766H}. According to this, large grains exposed to intense radiation fields are spun up to extreme suprathermal rotation rates. When the resulting centrifugal stress exceeds their tensile strength, they fragment into smaller grains \citep{2020ApJ...896...44L}. This process leads to a depletion of large, well-aligned grains. Since, RAT-A is less effective for smaller grains, the grain alignment efficiency and hence polarization percentage decrease.

To understand how magnetic field tangling contributes to the observed depolarization in the dense and high-temperature regions, we calculated the polarization angle dispersion function $S$ for our observed data (See  \citealp{2020A&A...641A..12P} for details regarding its calculation and noise bias correction). The values of $S$ quantifies the degree of local irregularity in the magnetic field orientation within the region. In addition, we evaluated the product 
$P \times S$, which serves as a diagnostic of the average grain alignment efficiency along the line of sight \citep{2025ApJ...981..128P}.

\begin{figure*}
\begin{subfigure}{\columnwidth}\hspace{0.5cm}
\includegraphics[width=0.8\columnwidth]{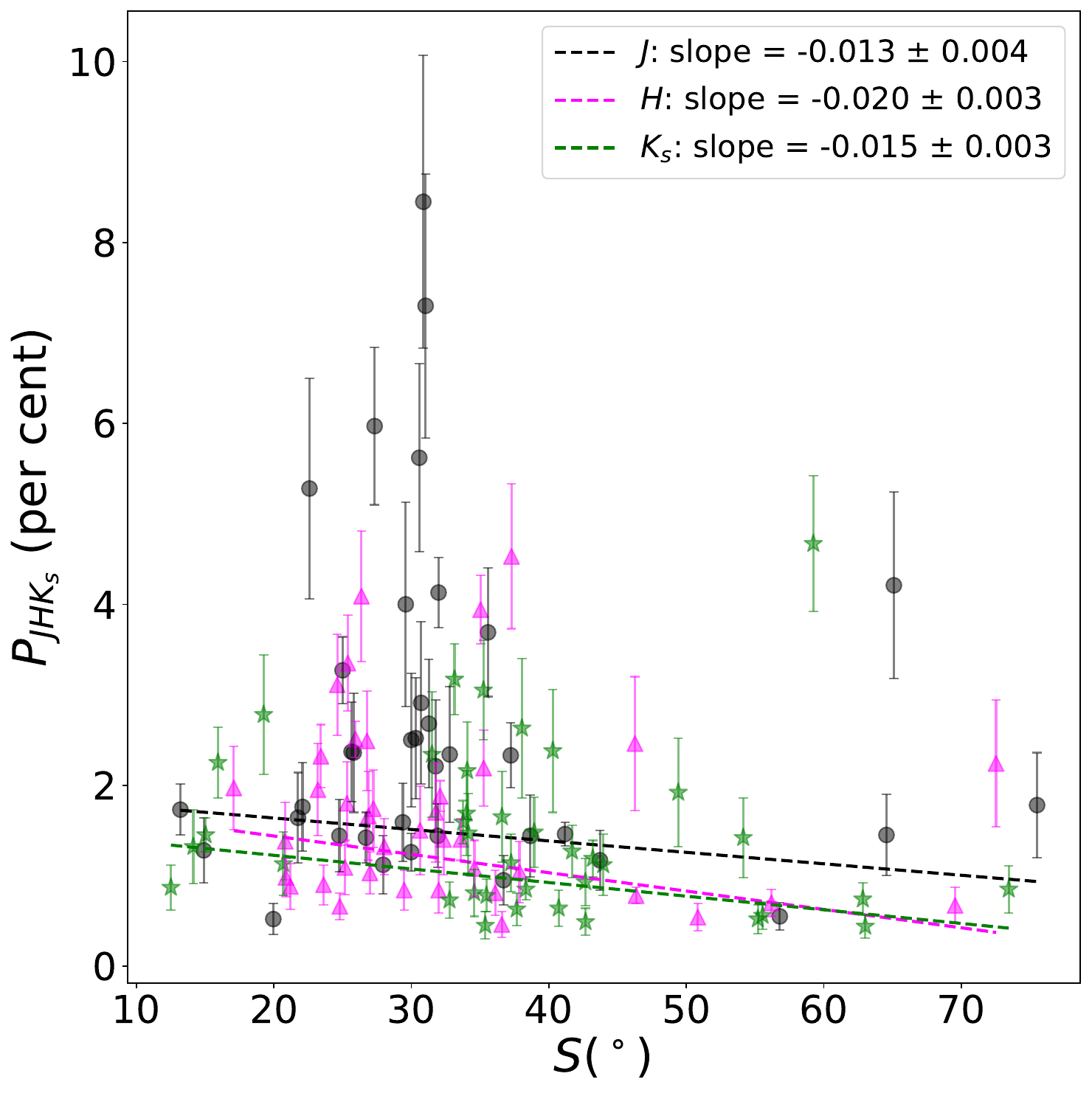}
\caption{}
\label{Fig:PvsS}
\end{subfigure}
\begin{subfigure}{\columnwidth}\hspace{0.5cm}
\includegraphics[width=0.8\columnwidth]{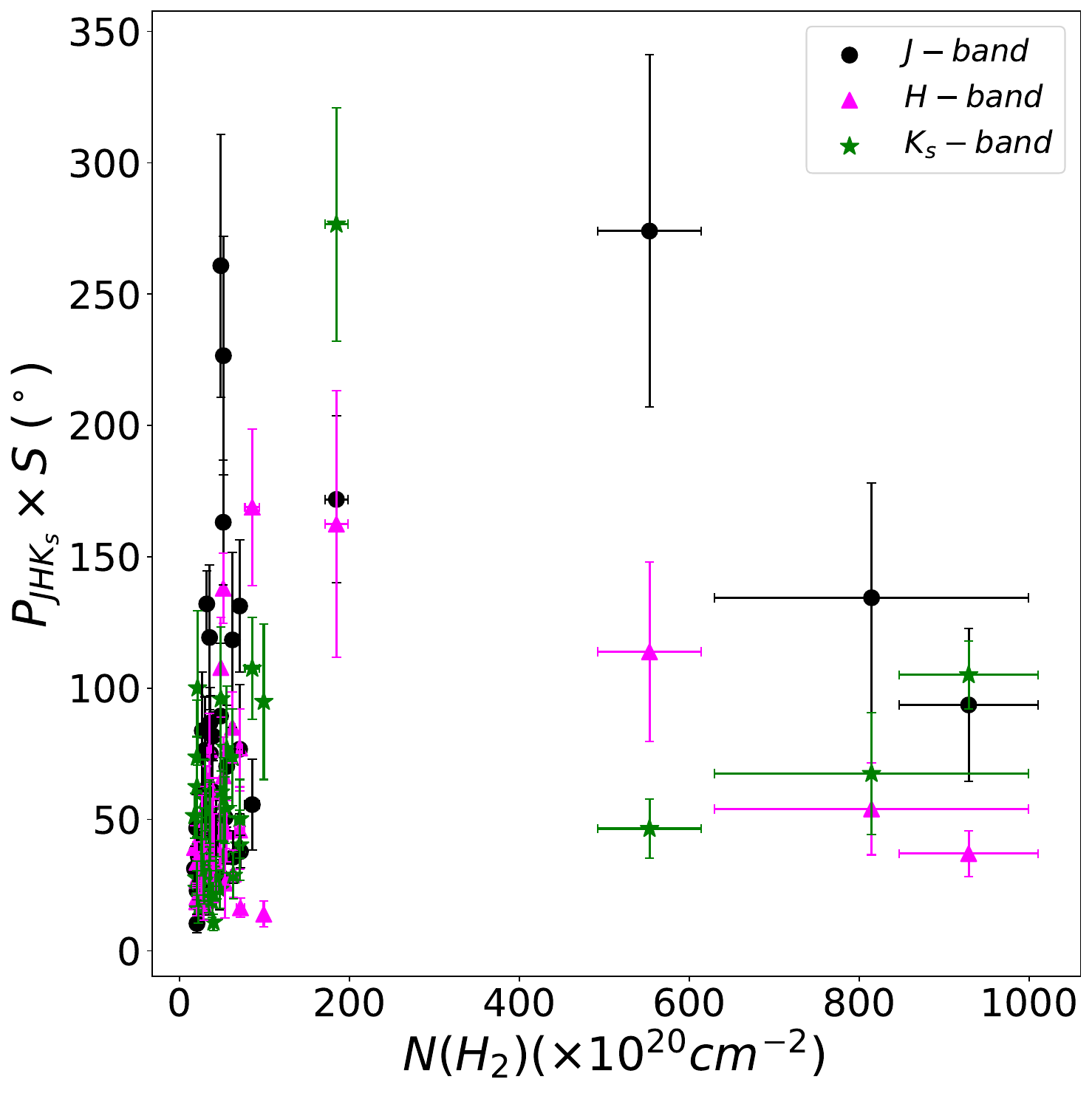}
\caption{}
\label{Fig:effvsnh2}
\end{subfigure}
\caption{(a) Variation of polarization percentage with the polarization angle dispersion function. The black circles, magenta triangles and green star symbols correspond to the $J$, $H$, and $K_s$ polarization, respectively. The dotted lines of corresponding colors represent the best-fit trends for each band. (b) Variations of the averaged grain alignment efficiency with the molecular hydrogen column density. The symbols have the same meaning as the Fig.~\ref{Fig:PvsS}}
\label{Fig:ga_eff}
\end{figure*}

Fig.~\ref{Fig:PvsS} illustrates the variation of $P_{JHK_s}$ as a function of $S$. The black circles, magenta triangles, and green star symbols correspond to the $J$, $H$, and $K_s$ polarization, respectively. The dotted lines of corresponding colors represent the best-fit trends for each band. The fitted lines are intended primarily to illustrate the average behavior of polarization with increasing $S$.  The derived slopes are $-0.013 \pm 0.004$ ($J$ band), $-0.020 \pm 0.003$ ($H$ band), and $-0.015 \pm 0.003$ ($K_s$ band). The fits use inverse-variance weighting based on the polarization uncertainties. The values of the fitted slopes suggest that polarization decreases modestly as the irregularity in the magnetic field orientation increases. Nevertheless, the small magnitudes of the slopes indicate that depolarization due to magnetic field tangling exists, but its effect is relatively small.

Fig.~\ref{Fig:effvsnh2} shows the variation of $P_{JHK_s} \times S$  as a function of $\mathrm N(\mathrm H_2)$. It uses the same color coding as Fig.~\ref{Fig:PvsS}. A comparison of Fig.~\ref{Fig:effvsnh2} and Fig.~\ref{Fig:cdens_temp_pol} demonstrates that the changes in grain alignment efficiency with $\mathrm{N}(\mathrm{H}_2)$ closely mirror the observed variations in the polarization with respect to $\mathrm{N}(\mathrm{H}_2)$. In other words, higher polarization corresponds with regions of elevated $P \times S$, while lower polarization is observed where $P \times S$ is reduced. This coherent behavior across the full column-density range along with the minimal change (or decrease) in polarization with $S$ indicates that the observed polarization/depolarization behavior in this region is predominantly governed by $P \times S$, i.e., the efficiency of dust grain alignment rather than the magnetic field tangling.

\subsection{Magnetic field orientation}
\label{sec:mag_field}

\begin{figure*}\hspace{1cm}
\includegraphics[width=1.8\columnwidth]{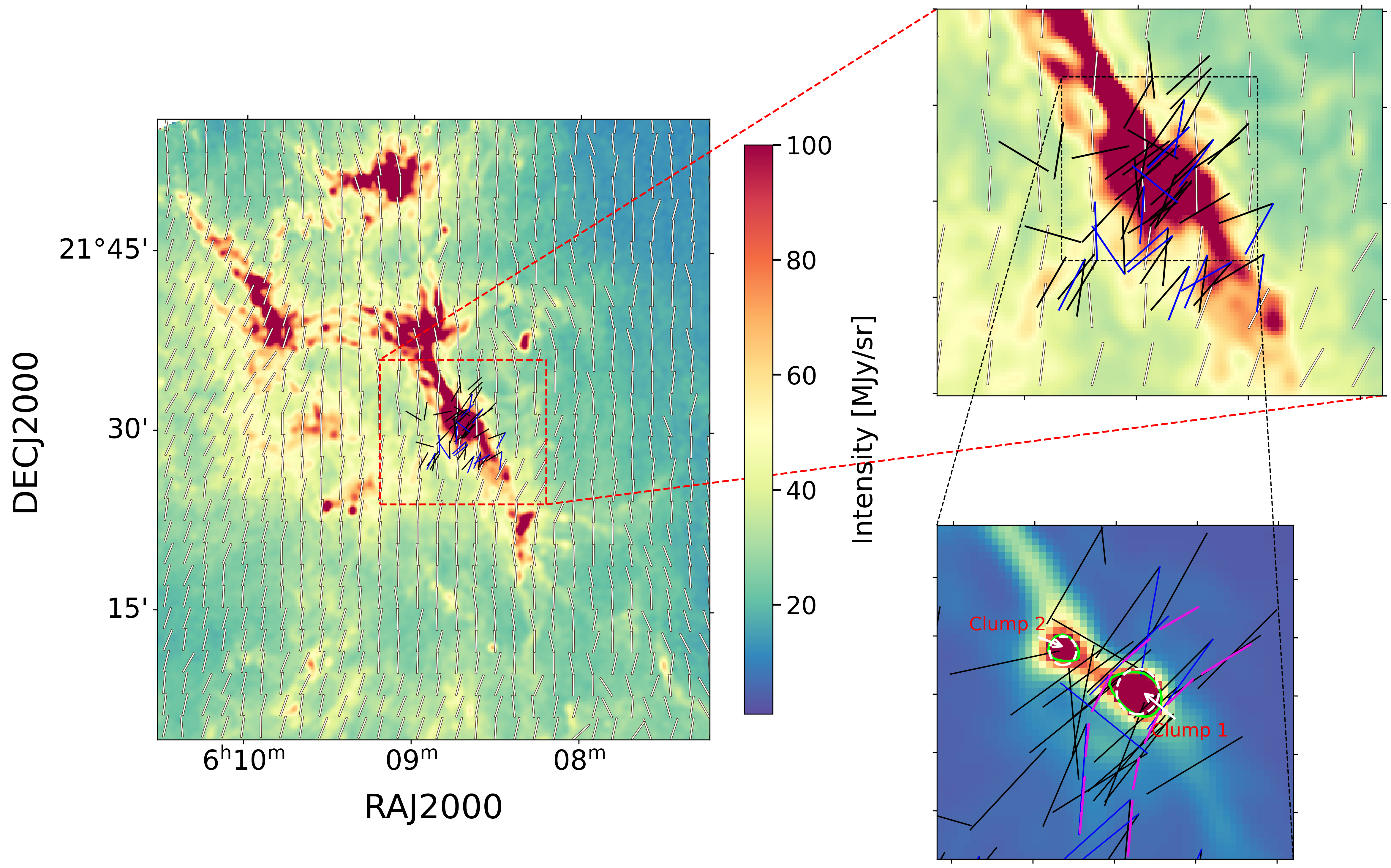}
\caption{A $50 \times 50$~arcmin$^2$ \textit{Herschel} SPIRE 500 $\mu$m dust continuum emmission map of the AFGL~6366S region overlaid with polarization vectors from the \textit{Planck}-HFI 353~GHz data (white) as well as the observed $H$-band (black) and $R_c$-band (blue) data from SIRPOL and AIMPOL, respectively. The top-right inset  provides a magnified view of the central cluster region. The bottom-right inset offers a further zoomed in view, highlighting two dense molecular clumps in the region, using green contours at a level of 1000 MJy/Sr. The clumps are approximated with circular geometries, indicated by white dashed circles. The magenta dashed curves trace the U-shaped morphology of the magnetic field.}
\label{Fig:planck_pol_map}
\end{figure*}

The position angle of the observed polarization is a well-established tracer of the magnetic field orientation in a region \citep{medhi_2007, Medhi_2008, Medhi_2010, Biswas_2024}. Hence, to understand the magnetic field structure towards the region of AFGL 6366S, we integrated the all-sky \textit{Planck}-HFI 353~GHz polarization data with our Optical and NIR polarization measurements. This facilitated a direct comparison between the large-scale Galactic magnetic field, traced by the \textit{Planck} data and the local field probed by our observations. Again, the stars showing intrinsic polarization signatures were excluded from the analysis. These may otherwise cause as a misinterpretation of the true field orientation \citep{Biswas_2024}.

Fig.~\ref{Fig:planck_pol_map} shows the \textit{Herschel} SPIRE 500 $\mu$m dust continuum emmission map of the AFGL~6366S region, across a $50 \times 50$~arcmin$^2$ field. It is overlaid with polarization vectors from the \textit{Planck}-HFI 353~GHz data (shown in white color). In addition, we superimposed the polarization vectors obtained from our $H$-band (black) and $R_c$-band (blue) polarimetric observations. All the polarization vectors are plotted with uniform lengths, to facilitate better visual comparison of their orientations. The \textit{Planck} vectors, with orientation $\theta_B$ (See Eq. \ref{Eq:planck_theta}), were interpolated from the original \texttt{HEALPix} grid onto a $1000 \times 1000$~pixel$^2$ R.A. - Dec. grid. 

The top-right inset of Fig.~\ref{Fig:planck_pol_map}  provides a magnified view of the central cluster region. It clearly shows that the local magnetic field orientation (traced by the blue and black vectors) deviates significantly from the large scale Galactic field (traced by the white vectors). These suggests that local turbulence and feedback from ongoing star formation are shaping the field morphology in the central cluster region. Furthermore, the local magnetic field also shows a large angular offset from the filament axis orientation ($\theta_{\rm fil} \sim 216^\circ$; see Sec. \ref{Sec: cdens_dtemp}). Such a configuration is consistent with the tendency of magnetic fields to exhibit near perpendicular offsets to the long axes of dense filaments \citep{2024MNRAS.528.1460R}.


We computed the average (inverse-variance–weighted) of the observed position angles in each wavelength band to represent the local magnetic field direction ($\theta_{B}^{env}$). Table~\ref{tab:mag_field_ori} lists the resulting $\theta_B^{env}$ values, for all observed stars and the subset member population, along with their angular offsets from the Galactic field direction (characterized by $\theta_{\rm GP} = 151.04^\circ$; see Sec. \ref{sec:result_1}). These offsets range from $0.26^\circ$ to $30.46^\circ$. The offsets between $\theta_{B}^{env}$ and the filament axis, denoted as $\theta_{\mathrm{offset}}^{fil}$ range from $\sim 34^\circ$–$73^\circ$. These are also listed in Table~\ref{tab:mag_field_ori}

The bottom-right inset of Fig.~\ref{Fig:planck_pol_map} offers a further zoomed in view, of the central region. Here, the magnetic field vectors appear to wrap around two dense cores, which are shown using the green contours at a level of 1000 MJy/sr. These are labeled as Clump 1 (centered approximately at R.A. = 06$^h$ 08$^m$ 40.0$^s$, Dec. = 21$^\circ$31$'$06$''$; radius $\approx$ 0.34 pc) and Clump 2 (R.A. = 06$^h$ 08$^m$ 45.6$^s$, Dec. = 21$^\circ$31$'$48$''$; radius $\approx$ 0.22 pc). The clump centres were identified as the positions of peak intensity within their respective contours. Their effective radii were estimated from the area (A) enclosed by the contour using $R_{\rm eff} = \sqrt{A/\pi}$. This calculation assumes a circular approximation for the clump geometry, illustrated by the white dashed circles. We also estimated the mean column densities of the clumps 1 and 2 as $(7.9 \pm 1.1) \times 10^{22}$ cm$^{-2}$  and $(4.3 \pm 0.5) \times 10^{22}$ cm$^{-2}$, respectively. These estimates were again derived by analyzing the column-density map using \href{https://sites.google.com/cfa.harvard.edu/saoimageds9/download}{SAOImage DS9}. 


It is evident from Figs. \ref{Fig:cdens_map} and \ref{Fig:planck_pol_map}, that the clumps 1 and 2, appear to remain embedded within the filamentary structure. However, their column densities, especially for the clump 1, are much higher than that of the filament. This may indicate that the clumps represent regions of locally enhanced mass accumulation. Their formation could be influenced by turbulence, magnetic fields, and gravitationally driven flows. The filament likely acts as a feeder structure, channeling material into the condensations. This enables them to evolve more rapidly than their surroundings. Further numerical modelling and magnetohydrodynamic (MHD) simulations are required to quantify this effect.

Simulations of globally collapsing magnetized molecular clouds indicate that gravity-driven mass inflows from filaments into embedded clumps or cores can deform the magnetic field by drawing the field lines inward with the gas. This mechanism naturally produces the characteristic U-shaped curvature of magnetic field lines across the filament axis \citep{2018MNRAS.480.2939G, 2019MNRAS.490.3061V, 2024MNRAS.528.1460R}. Our data exhibit indications of a similar configuration. It is illustrated by the magenta dashed curve in the bottom-right inset of Fig.~\ref{Fig:planck_pol_map}. From the figure, the axis of curvature of the field lines may be interpreted as a tracer of the dominant direction of mass inflow, suggesting that the flow is largely perpendicular to the local magnetic field direction. Definitive confirmation of these features, however, will require observations with significantly higher resolution and sensitivity.

\begin{table*}
\setlength{\tabcolsep}{6pt}
\centering
\caption{Weighted mean magnetic field orientations ($\theta_B^{\rm env}$) in different filters (all stars and members), with absolute offsets from the Galactic plane ($\theta_{\rm offset}$) and the filament ($\theta_{\rm offset}^{\rm fil}$). $\theta_{\rm GP} \sim 151^\circ$ represents the Galactic plane orientation and $\theta_{\rm fil} \sim 216^\circ$ represents the filament orientation.}
\label{tab:mag_field_ori}
\renewcommand{\arraystretch}{1.3}
\begin{tabular}{lcccc}
\hline
Filter & $\theta_B^{env}$ (All Stars) & $\theta_B^{env}$ (Members Only) & $\theta_\mathrm{offset} = |\theta_B^{env} - \theta_{\rm GP}|$ (All / Members) & $\theta_\mathrm{offset}^{\rm fil} = |\theta_B^{env} - \theta_{\rm fil}|$ (All / Members) \\
& \text{(°)} & \text{(°)} & \text{(°)} & \text{(°)} \\
\hline
$J$   & $154.73 \pm 6.46$  & $148.58 \pm 4.10$  & $3.69 / 2.46$ & $61.27 / 67.42$\\
$H$   & $149.38 \pm 6.34$  & $143.22 \pm 5.58$  & $1.66 / 7.82$ & $66.62 / 72.78$\\
$K_s$ & $172.13 \pm 6.99$  & $151.30 \pm 10.38$ & $21.09 / 0.26$ & $43.87 / 64.70$\\
$B$   & $175.67 \pm 7.09$  & $177.41 \pm 2.64$  & $24.63 / 26.37$ & $40.33 / 38.59$\\
$V$   & $169.78 \pm 7.22$  & $1.50 \pm 1.77$    & $18.74 / 30.46$ & $46.22 / 34.50$\\
$R_c$ & $155.77 \pm 8.12$  & $173.99 \pm 2.64$  & $4.73 / 22.95$ & $60.23 / 42.01$\\
$I_c$ & $160.06 \pm 9.13$  & $175.61 \pm 5.27$  & $9.02 / 24.57$ & $55.94 / 40.39$\\
\hline
\end{tabular}
\setlength{\leftskip}{-10pt}\\
\end{table*}

\subsection{Magnetic field strength and clump criticality}
\label{sec:mag_field_strength}

The Davis–Chandrasekhar–Fermi (DCF) method \citep{davis_1951, 1953ApJ...118..113C} estimates magnetic field strength from polarization angle dispersion and non-thermal gas motions. However, it is reliable only for angular dispersion $\lesssim 25^\circ$ \citep{2017MNRAS.464.2403S}, which is not the case for our data. Hence, we determine the POS magnetic field strength (B$_{\rm pos}$) within the region of our polarimetric observation (specifically, in the region of clumps 1 and 2) using the modified Chandrasekhar-Fermi (CF) relation \citep{2015ApJ...807....5F}. It is based on the structure function (SF) analysis technique \citep{2008ApJ...679..537F, 2009ApJ...696..567H}. 

The SF is defined as the mean squared difference between polarization angles measured at two positions separated by a distance $l$ \citep{2008ApJ...679..537F}. 

If the magnetic field in a region is assumed to be the sum of an ordered large-scale component, $B_o(x)$ with variation length scale $d$, and a turbulent component, $B_t(x)$ with correlation length scale $\delta$,  then for $\delta < l << d$, the SF can be approximated as \citep{2009ApJ...696..567H}:

\begin{equation}
	\langle \Delta(\phi)^2(l) \rangle_{\rm tot} \simeq b^2 + m^2l^2 + \sigma_M^2(l) 
	\label{Eq:ADF}
\end{equation}

where $\langle \Delta(\phi)^2(l) \rangle_{\rm tot}$ is the total observed polarization angle dispersion, $\sigma_M^2(l)$ represents the measurement uncertainty of SF as a function of $l$, $b^2$ denotes the constant turbulent contribution (the intercept of the curve after subtracting $\sigma_M^2(l)$) and $m^2l^2$ accounts for the ordered field component.

\begin{figure}\hspace{0.25cm}
\includegraphics[width=0.9\columnwidth]{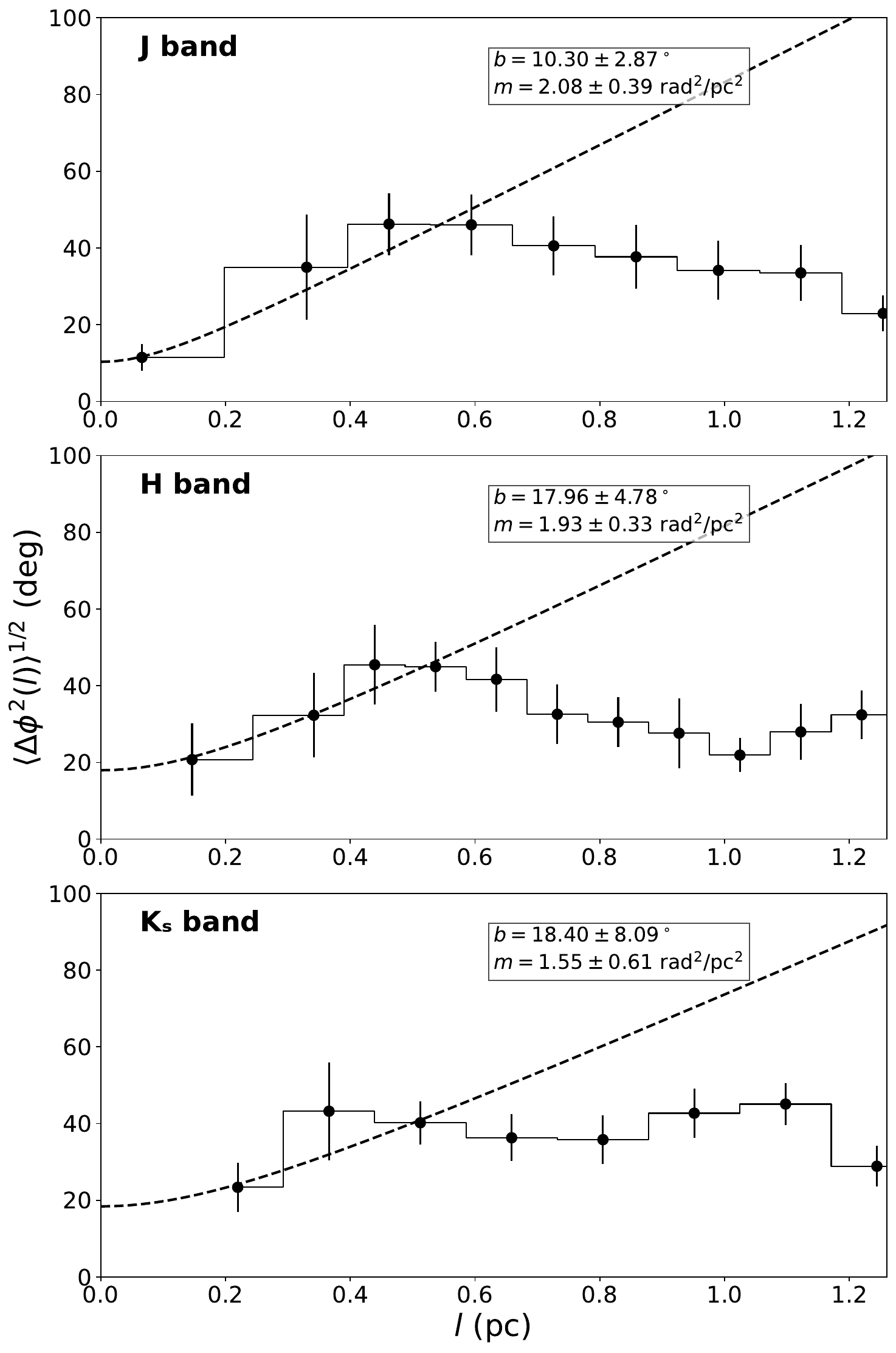}
\caption{Angular dispersion function (ADF) of polarization angles in the $J$, $H$, and $K_s$ bands as a function of projected separation for 40 ($J$ band), 42 ($H$ band), and 41 ($K_s$ band) stars observed toward AFGL~6366S. The dashed curves represent the best-fit models to the data for separations $< 0.6$~pc. The corresponding best-fit parameters, $b$ and $m$, are indicated in each panel.}
	\label{Fig:ADF}
\end{figure}

The square root of the SF is known as the angular dispersion function (ADF) \citep{2008ApJ...679..537F}. For the observed $JHK_s$ polarimetric data (excluding all probable intrinsic polarization sources), we estimated the ADF separately for each band and plotted it as a function of $l$ as shown in Fig.~\ref{Fig:ADF}. The Optical $BVR_cI_c$ data were excluded from this analysis due to limited statistics. The data points (black) in Fig.~\ref{Fig:ADF} represent the ADF values corrected for measurement uncertainties, i.e., $\sqrt{\langle \Delta\phi^2(l) \rangle_{\rm tot} - \sigma_M^2(l)}$. 

Subsequently, the parameters $b$ and $m$, were determined by fitting the observed ADF, for each wavelength band, to Eq. \ref{Eq:ADF}. The corresponding best-fit models are shown in Fig.~\ref{Fig:ADF} as black dashed curves. The minimum value of $l$ used in the fit were 0.07~pc for the $J$ band, 0.15~pc for the $H$ band and 0.22~pc for the $K_s$ band. Since, $\delta$ in molecular clouds is typically $\sim$1 mpc \citep{2009ApJ...696..567H, 2009ApJ...706.1504H}, the condition $\delta < l$ is thus satisfied. Furthermore, to ensure compliance with the condition $l \ll d$, only the first few ADF points corresponding to separations $<0.6$~pc were included, beyond which the ADF flattens, indicating saturation and loss of ordered field coherence.


The best-fit values of $b$ were found to be 
$10.30 \pm 2.87^{\circ}$ (in the $J$ band), 
$17.96 \pm 4.78^{\circ}$ ($H$ band), 
and $18.40 \pm 8.09^{\circ}$ ($K_s$ band). 
The corresponding best-fit slope parameters $m$ were 
$2.08 \pm 0.40~\mathrm{rad^2\,pc^{-2}}$ ($J$ band), 
$1.93 \pm 0.33~\mathrm{rad^2\,pc^{-2}}$ ($H$ band), 
and $1.55 \pm 0.61~\mathrm{rad^2\,pc^{-2}}$ ($K_s$ band). 

Finally, B$_{\rm pos}$ was estimated from the modified CF relation \citep{2015ApJ...807....5F}:

\begin{equation}
	B_{\mathrm{pos}} = 9.3 
	\left[ \frac{2n_{\mathrm{H_2}}}{\mathrm{cm^{-3}}} \right]^{1/2} 
	\left[ \frac{\Delta V}{\mathrm{km\,s^{-1}}} \right] 
	\left[ \frac{b}{1^{\circ}} \right]^{-1} \, \mu\mathrm{G}.
\end{equation}

Here, $n_{\mathrm{H_2}}$ represents the molecular hydrogen number density and $\Delta V$ denotes the velocity dispersion of the gas, obtained from spectral line width measurements. The applicability of this relation is restricted to b $\ll$ 2 \citep{2015ApJ...807....5F}.

\begin{table*}
\centering
\caption{Clump properties: radius ($R$), $H_2$ column ($N(H_2)$) and number ($n_{\rm H_2}$) densities, gas velocity dispersion ($\Delta V$), POS magnetic field ($\overline{B}_{\rm pos}$), and mass-to-flux ratio ($\lambda$).}
\begin{tabular}{lcccccc}
\hline
\hline
Clumps  & $R$ &$N(H_2)$ & $n_{\mathrm{H_2}}$ & $\Delta V$  & $\overline{B}_{\rm pos}$& $\lambda$ \\
&($\mathrm{cm}$)& ($\mathrm{cm^{-2}}$) &  ($\mathrm{cm^{-3}}$) & ($\mathrm{km\,s^{-1}}$) & ($\mu$G)& \\
\hline
Clump 1 & $1.05 \times 10^{18} (\approx 0.34 ~\rm pc)$& $(7.9 \pm 1.1) \times 10^{22}$ & $(3.76 \pm 0.52) \times10^{4}$ & $2.8$ &  $447.91 \pm 83.81$ & 1.34 \\
Clump 2 & $0.68 \times 10^{18} (\approx 0.22 ~\rm pc)$& $(4.3 \pm 0.5) \times 10^{22}$ & $(3.17 \pm 0.37) \times10^{4}$ & $2.7$ &  $396.66 \pm 73.64$ & 0.82 \\
\hline
\end{tabular}
\setlength{\leftskip}{-35pt}\\
\label{tab:cores_mag_parameters}
\end{table*}

Assuming a uniform spherical geometry, the number density for the clumps 1 and 2 were estimated using the relation $n_{\mathrm{H_2}}$ = $\mathrm N(\mathrm H_2)/2R$, where $R$ denotes the clump radius \citep{2005ApJ...622..346C}. The resulting values were $(3.76 \pm 0.52) \times 10^4$ cm$^{-3}$ and $(3.17 \pm 0.37) \times 10^4$ cm$^{-3}$, respectively. The $\Delta V$ values were adopted as 2.8 and 2.7 for clump 1 and 2, respectively, from the $^{13}\rm CO$ linewidth measurements reported by \cite{shimoikura_2013}. Accordingly, $B_{\mathrm{pos}}$ were estimated to be $693.36 \pm  198.85~ \mu G$ (for $J$-band), $397.60 \pm 109.42 ~ \mu G$ (for $H$-band) and $388.11 \pm 172.67 ~ \mu G$ (for $K_s$-band) in clump 1. For clump 2, $B_{\mathrm{pos}}$ were estimated as $613.90 \pm 174.58~ \mu G$ (for $J$-band), $352.03 \pm 96.00 ~ \mu G$ (for $H$-band) and $343.64 \pm 152.35 ~ \mu G$ (for $K_s$-band). Uncertainties in $B_{\mathrm{pos}}$ were calculated by propagating the errors in both $n_{\mathrm{H_2}}$ and $b$. Uncertainties in $\Delta V$ were not available. Thus, the reported error bars should be regarded as lower limits on the true uncertainties. The inverse variance weighted mean of magnetic field strength ($\overline{B}_{\mathrm{pos}}$), across the three bands were estimated as: 447.91 $\pm$ 83.81 $~ \mu G$ and 396.66 $\pm$ 73.64 $~ \mu G$, for clump 1 and clump 2, respectively. The derived $\overline{B}_{\mathrm{pos}}$ values and other physical parameters of the two clumps are summarized in Table \ref{tab:cores_mag_parameters}. As a caveat, the $\overline{B}_{\mathrm{pos}}$ estimates should be regarded as approximate, as the modified CF method carries order-unity systematic uncertainties including the effects from geometric factors, line-of-sight averaging and finite angular resolution \citep{2009ApJ...696..567H, 2009ApJ...706.1504H}.

\subsection{Mass-to-flux ratio}

The relative influence of magnetic fields in opposing self-gravity can be evaluated using the mass-to-flux ratio, which quantifies whether the magnetic field is sufficiently strong to support a cloud against gravitational collapse \citep{1976ApJ...210..326M}. This ratio is expressed through the dimensionless parameter $\lambda$ \citep{1978PASJ...30..671N, 2004mim..proc..123C}, given by

\begin{equation}
	\lambda = 7.6 \times 10^{-21} \frac{N_{\mathrm{H_2}}}{B_{\mathrm{pos}}},
\end{equation}

If $\lambda < 1$, the region is magnetically subcritical, implying that the magnetic field provides sufficient support against gravitational collapse. Conversely, for $\lambda > 1$, the region is magnetically supercritical, where gravity dominates over magnetic support.

For clumps 1 and 2, the derived $\lambda$ values are 1.34 and 0.82, respectively. This indicates that clump 1 is magnetically supercritical, whereas clump 2 is magnetically subcritical. Thus, it suggests that clump 1 is in a more advanced stage of contraction. 

\subsection{Alfv\'en Mach Number}

The relative importance of turbulence and magnetic fields in a medium can be evaluated through the Alfv\'en Mach Number ($\mathcal{M}_A$). It is expressed as \citep{1999ApJ...514L.121C}:

 \begin{equation}
     \mathcal{M}_A = \frac{\sigma_{\rm NT}}{V_A}
 \end{equation}

  where, $V_A$ = $\overline{B}_{\mathrm{pos}}/\sqrt{4\pi\rho}$ represents the Alfv\'en speed. The parameter $\rho$ = $n_{\rm H_2}\mu m_H$ (where, $m_{\rm H}$ = hydrogen mass; $\mu$ = atomic weight of the observed molecule) denotes the mass density. The derived mass densities for clumps 1 and 2 are $\rho \sim 1.76 \times 10^{-19}$ and $1.49 \times 10^{-19}$ g cm$^{-3}$, respectively. The corresponding Alfv\'en speeds are $V_A \sim 3.01$ and 2.90 km s$^{-1}$. 
  
  The parameter $\sigma_{\rm NT}$ is the non-thermal velocity dispersion and is expressed as \citep{2022AJ....164..175C}:

  \begin{equation}
     \sigma_{\rm NT} = \sqrt{\left(\frac{\Delta V}{\sqrt{8~ln~2}}\right)^2 -\frac{k_BT_k}{\mu m_H}}
 \end{equation}
 
 Here, $k_{\rm B}$ is the Boltzmann constant, $T_{\rm k}$ is the kinetic gas temperature.

Assuming the clumps are in local thermodynamic equilibrium (LTE), the kinetic gas temperature can be approximated by the excitation temperature ($T_{\rm ex}$) \citep{2024MNRAS.528.2199R}. From \citet{shimoikura_2013}, we adopt $T_{\rm ex} \approx T_{\rm k} = 17.4~\mathrm{K}$ for both clumps. Thus, the non-thermal velocity dispersions are calculated to be $\sigma_{\rm NT} \sim$ 1.19 and 1.14 km s$^{-1}$. These values give Alfv\'en Mach Numbers as $\mathcal{M}_A$ $\sim$ 0.395 and 0.393 for clump 1 and 2, respectively.

The values of $\mathcal{M}_A < 1$ indicate that both clumps are in a sub-Alfv\'enic state. Under these conditions, the magnetic field provides the dominant dynamical influence, guiding and channeling the gas along the field lines. Turbulent motions become secondary and contribute less to shaping the overall structure.

\section{DISCUSSION}
\label{sec:discuss}

A consistent interpretation of our results requires clarifying how the various stellar sub-samples relate to one another. The study combines multi-band polarimetry (Optical and NIR) and NIR photometry, consequently, the number of stars contributing to each diagnostic naturally differs. These variations arise from wavelength-dependent detection limits, \(\mathrm{SNR}\) requirements, and the specific criteria of each analysis. 

For example, the Optical Serkowski-law fits rely on stars with high-\(\mathrm{SNR}\) Optical polarization detections in at least three filters, resulting in a sample of \(19\) stars. The NIR wavelength-dependence analysis uses \(34\) stars that have good \(\mathrm{SNR}\) in all three NIR bands (\(J\), \(H\), and \(K_s\)) and exhibit the expected decrease in polarization from \(J\) to \(K_s\). The NIR PCD is constructed from \(53\) stars with good \(\mathrm{SNR}\) in the $JHK_s$ polarimetric data and reliable 2MASS photometry. The polarization-efficiency diagrams require both \(\lambda_{\rm max}\) and \(E(B-V)\), which limits that sample to \(15\) stars. 

A total of 22 stars show indications of intrinsic polarization, identified through a combination of NIR and Optical polarimetric diagnostics. These stars are also excluded from subsequent analyses of grain alignment and magnetic-field morphology to ensure that the inferred results represent purely interstellar contributions. Consequently, these grain alignment and magnetic-field analyses are based on the 40, 42 and 41 stars with $J,H ~\rm and ~K_s$ band polarimetric data, respectively. The complementary Optical polarimetric data are not used in these analyses owing to their limited sample size and their greater susceptibility to dust extinction.

Importantly, all sub-samples used across the various analyses are drawn from the same parent set of 64 stars subjected to NIR polarimetric analysis.

\section{Summary}
\label{sec:summ}

In this study, we conducted NIR ($JHK_{s}$) and Optical ($BVR_cI_c$) linear polarimetric observations towards the young, partially embedded cluster AFGL~6366S. We also used numerous multi-wavelength archival datasets. We carried out a detailed analysis of dust properties, grain-alignment mechanisms,  magnetic-field morphology, criticality of the star forming clumps and overall integration of these factors across the target region. The principal findings are summarized as follows:

\begin{enumerate}
\renewcommand{\labelenumi}{(\roman{enumi})}
	\item The observed polarization toward AFGL~6366S ranges from 0.44-10.3 per cent in NIR  and 0.16-11.22 per cent in Optical bands. The position angle ranges $1^\circ - 179^\circ$ in both NIR and Optical bands. A total of 22 stars show probable signatures of intrinsic polarization.

	\item The central cluster region exhibits the highest values of
	\(N(\mathrm{H}_2) \approx 3.4 \times 10^{23}\,\mathrm{cm}^{-2}\) and 
	\(T_d \approx 28.8\,\mathrm{K}\), likely reflecting heating by embedded YSOs. A distinct filamentary structure is also apparent in this region. It is characterized by a mean column density of
	\((4.1 \pm 0.3) \times 10^{22}\,\mathrm{cm}^{-2}\).

	\item The polarization exhibits an initial increase with rising column density and dust temperature, followed by a clear depolarization trend at higher values. The initial positive correlation is consistent with the RAT-A theory. The depolarization trend in the densest and warmest zones are likely dominated by RAT-D mechanism with a small influence from magnetic-field tangling.
	
    \item Within the central cluster region, the local magnetic field exhibits large angular offset ($\sim~ 34^\circ - 73^\circ$, across different wavelength bands) from the filamentary axis. It also exhibits pronounced departures (spanning $0.26^\circ$ to $30.46^\circ$) from the large-scale Galactic field. The field morphology further wraps around two dense cores, clumps 1 and 2, which have radii $\sim$ 0.35 and 0.22 pc and $N(H_2)$ $\sim$ $(7.9 \pm 1.1) \times 10^{22}$ cm$^{-2}$  and $(4.3 \pm 0.5) \times 10^{22}$ cm$^{-2}$, respectively. Evidences of the characteristic U-shaped curvature of magnetic field lines across the filament axis were also found. This was possibly created due to the gravitationally driven mass flows, largely perpendicular to the local magnetic field, from filaments into the embedded clumps.

    \item The POS magnetic field strengths for clump~1 and clump~2 are 
    \(\overline{B}_{\mathrm{pos}} = 447.91 \pm 83.81~\mu\mathrm{G}\) and 
    \(396.66 \pm 73.64~\mu\mathrm{G}\), respectively. 
    The corresponding mass-to-flux ratios, \(\lambda = 1.34\) and \(0.82\), indicates that clump 1 is magnetically supercritical, whereas clump 2 is magnetically subcritical. Thus, it suggests that clump 1 is in a more advanced stage of contraction.

    \item The $\mathcal{M}_A$ values of $\sim$ 0.395 and 0.393, for the two clumps, indicate that both the clumps are in a sub-Alfv\'enic state. Hence, the magnetic fields play dominant role in the gas dynamics of the region, rather than the turbulence.

\end{enumerate}

Together, these results present a cohesive picture of the physical processes shaping AFGL~6366S region. They demonstrate how magnetic fields interact with the evolving density structure, guiding the interplay between gravitational forces and turbulence, and ultimately regulating the emergence and growth of star-forming condensations within the region.

\section*{Acknowledgments}

The authors thank the anonymous referee for their constructive comments that significantly improved this work. This research has made use of the VizieR catalogue access tool, CDS, Strasbourg, France; Aladin sky atlas developed at CDS, Strasbourg Observatory, France;  SIMBAD database, CDS, Strasbourg Astronomical Observatory, France;  \textit{Herschel} Science Archive; Digitized Sky Survey (DSS), which was produced at the Space Telescope Science Institute under the US Government grant NAG W-2166, NASA’s Astrophysics Data System. The IRSF project is a collaboration between Nagoya
University and the South African Astronomical Observatory
(SAAO) supported by Grants-in-Aid for Scientific Research on
Priority Areas (A; Nos. 10147207 and 10147214) and Optical
\& Near-Infrared Astronomy Inter-University Cooperation
Program, from the Ministry of Education, Culture, Sports,
Science and Technology (MEXT) of Japan and the National
Research Foundation (NRF) of South Africa.
Facility: IRSF;Software: IRAF.
S.B. highly acknowledges the Department of Science and 
Technology (DST), Govt. of India for providing DST INSPIRE 
fellowship vide grant no. IF220155. M.T. is supported by JSPS KAKENHI grant No. 24H00242. J.K. is supported by JSPS KAKENHI grant No. 24K07086. S.B. and B.J.M. acknowledge IUCAA, Pune, for providing access to the Pegasus High Performance Computing facility. The authors thank Eswaraiah Chakali, Archana Soam, Ekta Sharma, Saurabh Sharma, and Bhaskarjyoti Barman for valuable suggestions. B.J.M. would like to thank Orchid for her support.

\bibliography{AFGL2_refer}{}
\bibliographystyle{aasjournalv7}



\end{document}